 \documentclass[oneside,12pt]{Classes/CUEDthesisPSnPDF}

\ifpdf
    \pdfcatalog { /PageMode (/UseOutlines)
                  /OpenAction (fitbh)  }
\fi

\title{Safeguarding E-Commerce against Advisor Cheating Behaviors: Towards More Robust Trust Models for Handling Unfair Ratings}

\ifpdf
  \author{\href{mailto:y080077@e.ntu.edu.sg}{Lizi Zhang}}
  \collegeordept{\href{http://http://sce.ntu.edu.sg/Pages/Home.aspx}{School of Computer Engineering}}
  \university{\href{http://www.ntu.edu.sg}{Nanyang Technological University}}
  \crest{\includegraphics[width=30mm]{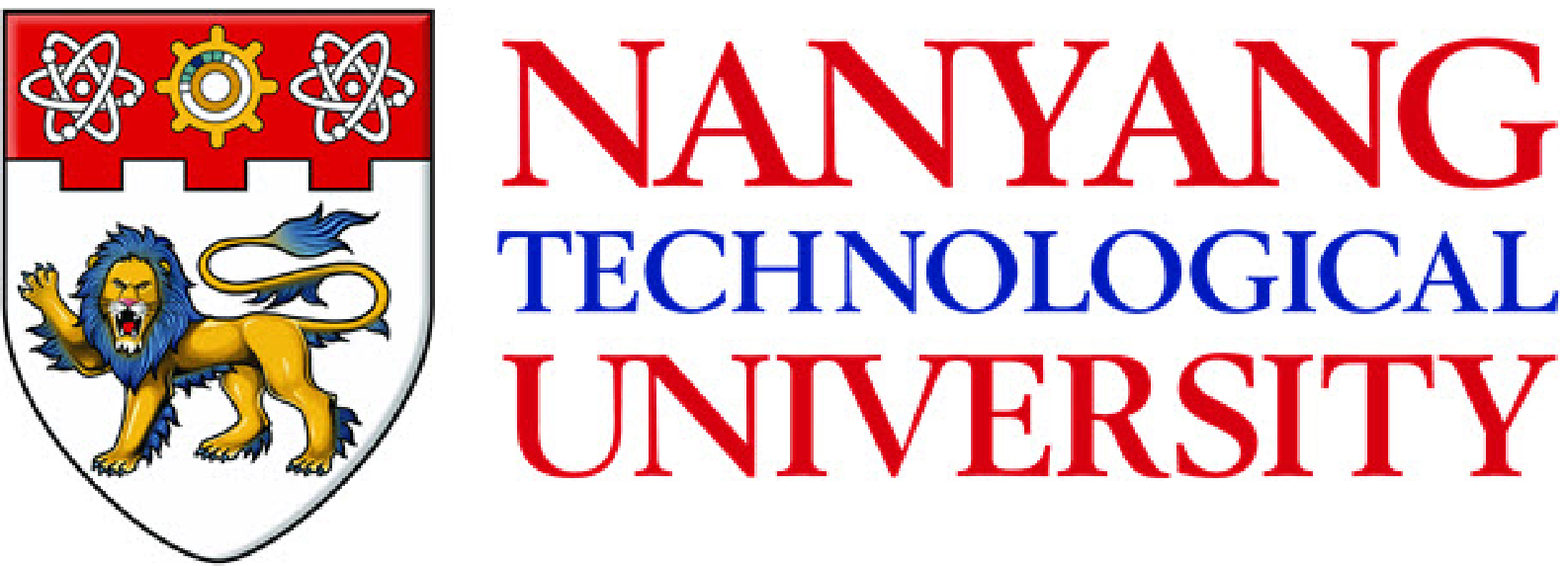}}
\else
  \author{Lizi Zhang}
  \collegeordept{School of Computer Engineering}
  \university{Nanyang Technological University}
  \crest{\includegraphics[width=70mm]{NTU_logo}}
\fi
\degree{Bachelor of Engineering (Computer Science)}
\degreedate{March 2012}

\usepackage{StyleFiles/watermark}

\begin{document}




\maketitle

\setcounter{secnumdepth}{3}
\setcounter{tocdepth}{3}

\frontmatter 
\pagenumbering{roman}


\begin{abstracts}        

In electronic marketplaces, after each transaction buyers will rate the products provided by the sellers. To decide the most trustworthy sellers to transact with, buyers rely on trust models to leverage these ratings to evaluate the reputation of sellers. Although the high effectiveness of different trust models for handling unfair ratings have been claimed by their designers, recently it is argued that these models are vulnerable to more intelligent attacks, and there is an urgent demand that the robustness of the existing trust models has to be evaluated in a more comprehensive way. In this work, we classify the existing trust models into two broad categories and propose an extendable e-marketplace testbed to evaluate their robustness against different unfair rating attacks comprehensively. On top of highlighting the robustness of the existing trust models for handling unfair ratings is far from what they were claimed to be, we further propose and validate a novel combination mechanism for the existing trust models, Discount-then-Filter, to notably enhance their robustness against the investigated attacks.

\end{abstracts}




\begin{acknowledgements}      

I would like to express my gratitude to the following people and organizations for their great supports and contributions in my Final Year Project:

\begin{itemize}
  \item \textbf{Dr. Ng Wee Keong}, my supervisor, for his insightful advice, constant help, invaluable encouragement and sharing of his knowledge. In 2010, inspired by my seniors' success, I became determined to do research as an undergraduate and contacted him to seek his guidance in my research. Since then, Dr. Ng has been supervising my research, and teaching me how to think deeply and present work clearly and logically. With his invaluable guidance, I attained great achievements in my research and published three research papers in international conferences during my undergraduate studies.
  \item \textbf{Dr. Zhang Jie}, my co-supervisor, for his utmost support throughout this project. His ideas and guidance kept this project in the right heading and his enthusiasm towards research always motivated me to move forward and further, especially during difficult times.
  \item \textbf{Mr. Jiang Siwei}, a Ph.D candidate at School of Computer Engineering, for his kind comments and help during the project.
  \item \textbf{all my peers, seniors and mentors at the Center for Computational Intelligence}, for cheering me up all the time.
  \item \textbf{Nanyang Technological University}, for the funding support for this project under the Final Year Project---Undergraduate Research Experience on CAmpus (FYP---URECA) program.
\end{itemize}

Last but not least, I would like to thank all my friends for their thoughtful support and encouragement during the project.

Without any of them, this work would not be possible.

Thank you all.

\end{acknowledgements}




\begin{dedication} 

I would like to dedicate this Final Year Project Report to my loving parents.

\end{dedication}



\tableofcontents
\listoftables
\listoffigures
\printnomenclature  

\mainmatter 
\chapter{Introduction}
\label{section:introduction}
\ifpdf
    \graphicspath{{Introduction/IntroductionFigs/PNG/}{Introduction/IntroductionFigs/PDF/}{Introduction/IntroductionFigs/}}
\else
    \graphicspath{{Introduction/IntroductionFigs/EPS/}{Introduction/IntroductionFigs/}}
\fi

\section{The Electronic Marketplace Environment}
\begin{figure}[t!]
\centering
\includegraphics[scale=0.55]{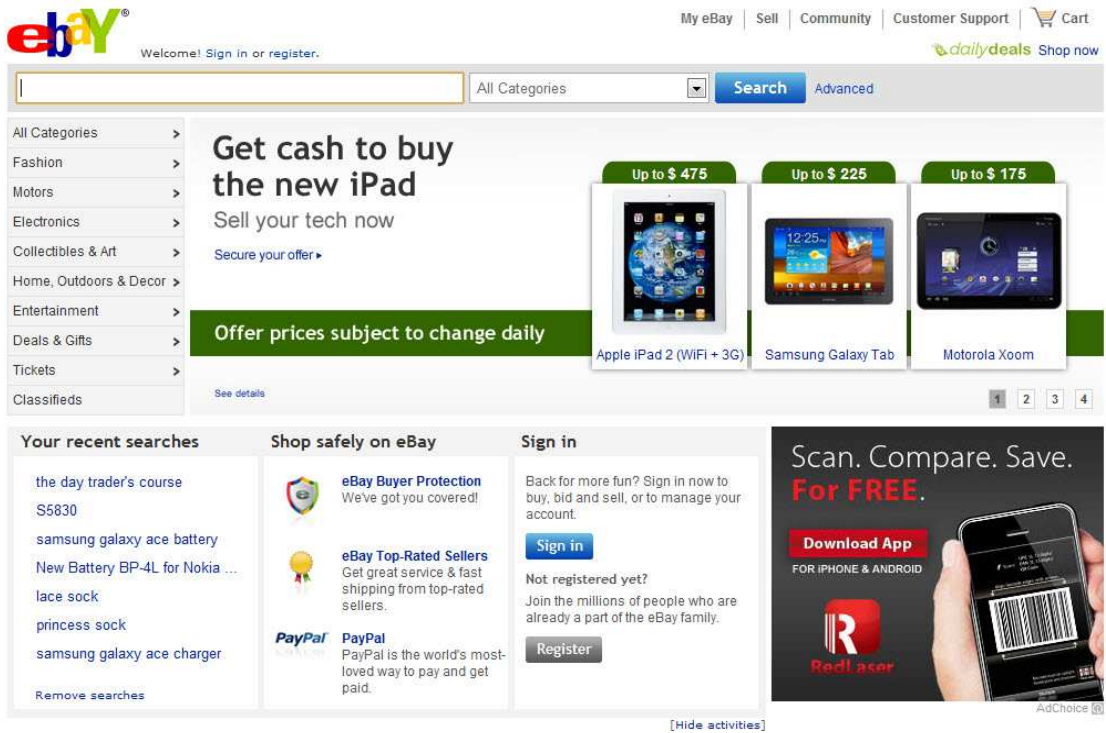}
\caption{Online shopping on eBay}
\label{fig:ebay_homepage}
\end{figure}

\markboth{\MakeUppercase{\thechapter. Introduction }}{\thechapter. Introduction}
Nowadays, electronic marketplaces, such as eBay (Fig.~\ref{fig:ebay_homepage}), have greatly facilitated the transaction processes among different people.
As the enterprise of electronic commerce becomes increasingly popular, worldwide, one challenge that arises is to ensure that organizations participating in e-commerce have
sufficient trust in order to bring their businesses on-line (\cite{zhang2006trusting}).
This is because, unlike traditional face-to-face transaction experiences, it is hardly possible for buyers to evaluate the products provided by sellers before they decide whether to buy from a potential seller.

\begin{figure}[t!]
\centering
\includegraphics[scale=0.6]{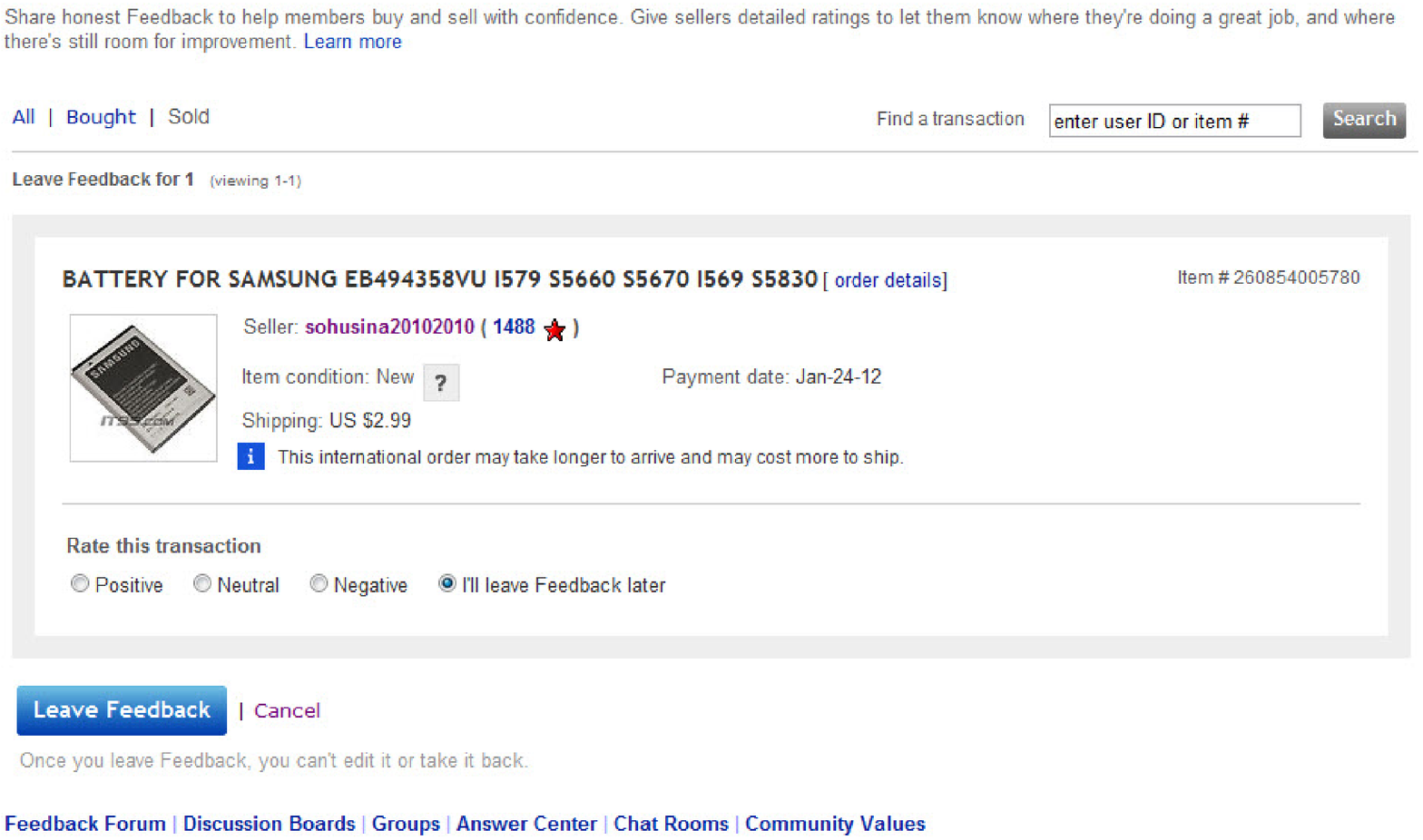}
\caption{Rating the seller after the transaction on eBay}
\label{fig:ebay_voting}
\end{figure}

\begin{figure}[t!]
\centering
\includegraphics[scale=0.65]{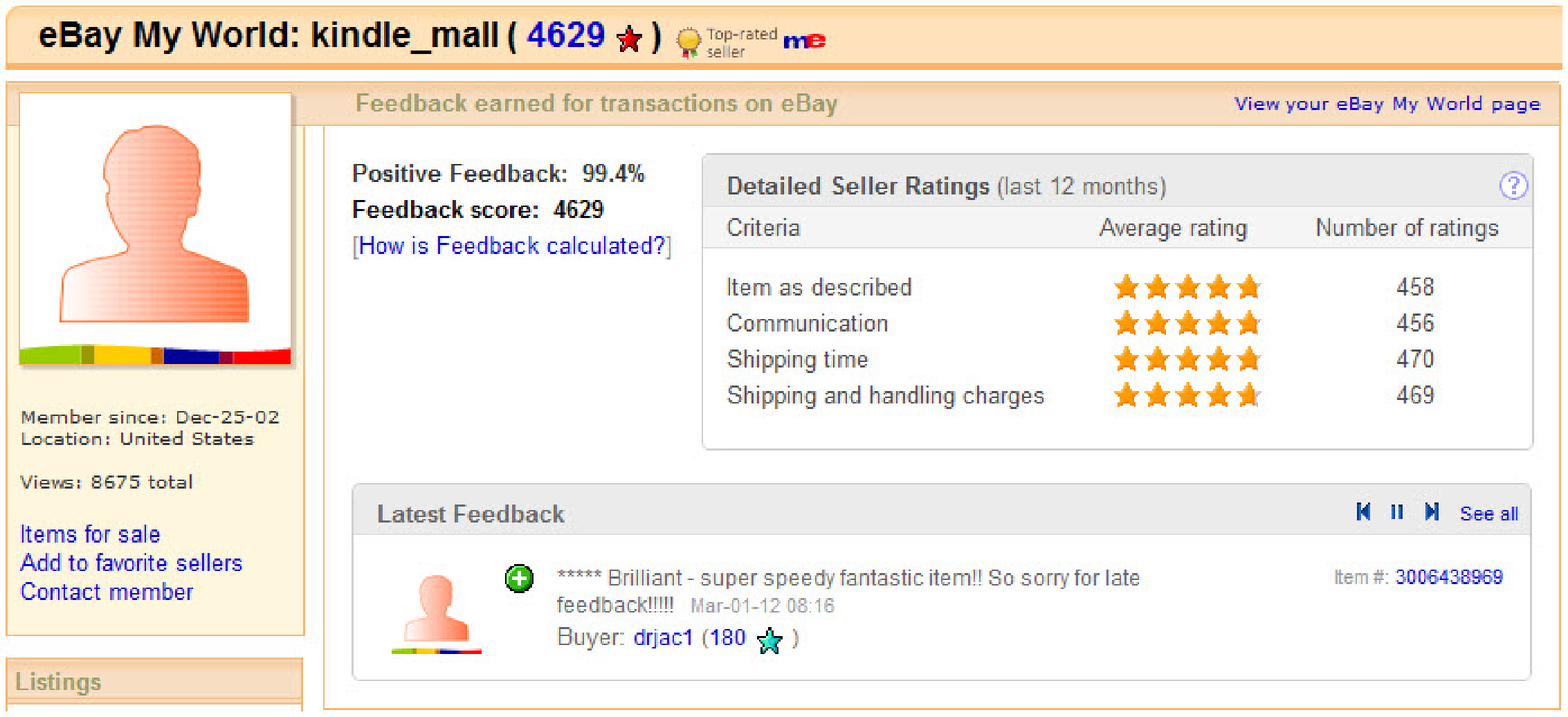}
\caption{Current seller reputation evaluation on eBay}
\label{fig:ebay_reputation}
\end{figure}

Current e-commerce systems like eBay, allow buyers to rate their sellers according to the quality of their delivered products after each transaction is completed (Fig.~\ref{fig:ebay_voting}).
In order to assist both individual buyers and business organizations in conducting both B2B and B2C e-commerce,
researchers in artificial intelligence have been designing intelligent agents to perform the tasks of buying or selling, on
behalf of their human clients (\cite{zhang2006trusting}).

In the context of the multiagent-based e-marketplace, when a buyer agent evaluates the reputation of a potential seller agent, he may need to ask for other buyer agents' opinions (advisors agents' ratings) towards that seller agent (Fig.~\ref{fig:ebay_reputation}). We define the following terms discussed in the remaining report:
\begin{itemize}
    \item \emph{Honest seller}: A seller that delivers his product as specified in the contract.
    \item \emph{Dishonest seller}: A seller that does not deliver his product as specified in the contract.
    \item \emph{Reputation}: A value calculated by trust models to indicate whether a seller will behave honestly in the future: the higher reputation, the higher probability that the seller will behave honestly.
    \item \emph{Positive rating}: A rating given by a buyer/advisor to a seller indicating a seller is an honest seller.
    \item \emph{Negative rating}: A rating given by a buyer/advisor to a seller indicating a seller is a dishonest seller.
    \item \emph{Honest buyer/advisor}: A buyer that always provides positive ratings to honest sellers or negative ratings to dishonest sellers.
    \item \emph{Dishonest buyer/advisor or Attacker}: A buyer that provides negative ratings to honest sellers or positive ratings to dishonest sellers. Exception: some special attacker (\emph{e.g.} Camouflage Attacker) may strategically behave like an honest buyer.
    \item \emph{Trust or Trustworthiness}: A value calculated by trust models to indicate whether an advisor is honest or not: the higher trustworthiness, the higher probability that the advisor is honest.
\end{itemize}

Notice that, generally, the terms \emph{reputation}, \emph{trust} and \emph{trustworthiness} are used interchangeably in many works. To avoid confusion, in this report we use them to model behaviors of sellers and buyers/advisors separately. In addition,
when a buyer evaluates a seller's reputation, other buyers become that buyer's advisors: a buyer $B_x$ seeks advice (in the form of ratings in  e-marketplaces) from his advisors $\{B_i | i \neq x\}$ who have transaction experience with the seller $S_y$ in the evaluation of $S_y$'s reputation. The terms \emph{advisor} and \emph{buyer} are used interchangeably in this report.

\section{Cheating Behaviors in Electronic Marketplaces}
Cheating behaviors from sellers, such as not performing the due obligations according to the transaction contract, are still possible to be sanctioned by law if trust models fail to take effect. However, advisors' cheating behaviors, especially providing \emph{unfair ratings} to sellers, are more difficult to be dealt with.

Dellarocas distinguished unfair ratings as unfairly high ratings (``ballot stuffing") and unfairly low ratings (``bad-mouthing") (\cite{cdellarocas00}).
Advisors may collude with certain sellers to boost their reputation by providing unfairly positive ratings while bad-mouthing their competitors' reputation with unfairly negative ratings. An example is that three colluded men positively rated each other several times and later sold a fake painting for a very high price (\cite{jzhang06}).

\section{Trust Models for Handling Unfair Ratings}
Trust has become a common and important issue since Web 2.0. Researchers studied trust and assisted people in choosing trustworthy online users in the context of various domains (\cite{josang2007survey}), such as forums (\cite{zhang2011intrank} and \cite{zhang2011influence}).

To address the above challenge emerging in the context of e-commerce, researchers in the multiagent-based e-marketplace have designed various \emph{trust models} (\emph{a.k.a.}, \emph{reputation systems} or \emph{trust and reputation systems}) to handle unfair ratings to assist buyers to evaluate the reputation of sellers more accurately. However, recently it was argued that the robustness analysis of these trust models had been mostly done through simple simulated scenarios implemented by the model designers themselves, and this cannot be considered as reliable evidence for how these systems would perform in a realistic environment (\cite{ajosang09}).

If a trust model is not \emph{robust against}, or \emph{vulnerable to}, certain unfair rating attack, mostly it will inaccurately compute a dishonest seller's reputation higher than that of an honest seller; thus, it will suggest honest buyers to transact with a dishonest seller, and sellers can gain higher transaction volumes by behaving dishonestly. If such dishonest behaviors---unfair ratings were encouraged and thus growing without being sanctioned in the e-marketplace, none of B2B and B2C e-commerce would survive. Therefore, there is an urgent demand to evaluate the robustness of the existing trust models under more comprehensive unfair rating attack environment before deploying them in the real market.

The ``Agent Reputation and Trust Testbed (ART)" (\cite{kFullam05}) is an example of a testbed that has been specified and
implemented by an international group of researchers. However, it is currently not flexible
enough for carrying out realistic simulations and robustness evaluations for many of the
proposed trust models (\cite{ajosang09}).

\section{Our Contributions}
In this work, we selected and investigated four well-known existing trust models (BRS, iCLUB, TRAVOS and Personalized) and six unfair rating attack strategies (Constant, Camouflage, Whitewashing, Sybil, Sybil Camouflage, and Sybil Whitewashing Attack). We classified these trust models into two broad categories: \emph{Filtering-based} and \emph{Discounting-based}, and proposed an extendable e-marketplace testbed to evaluate their robustness against different attacks comprehensively and comparatively. To the best of our knowledge, we for the first time experimentally substantiate the presence of their multiple vulnerabilities under the investigated unfair rating attacks.

On top of highlighting the robustness of the existing trust models is far from what they were claimed to be---none of the investigated single trust model is robust against all the six investigated attacks, we further proposed and validated a novel combination approach, \emph{Discount-then-Filter}, for the existing trust models. This combination notably enhanced their robustness against all the attacks: our experiments show most of Discount-then-Filter combined trust models are robust against all the six investigated attacks. Equipped with such combined trust models, e-commerce can be better safeguarded against unfair ratings---the advisor cheating behaviors.

\section{Report Organization}
The rest of the report is organized as follows. In Chapter~\ref{section:Relatedwork} we
consider related work and describe our investigated attack strategies and trust models. Chapter~\ref{section:Evaluationmethod} is about the e-marketplace testbed and the evaluation metric used in our experiments. Based on the experimental results, we compare and analyze the robustness of all the single trust models against each investigated attack in Chapter~\ref{section:RobustnessOfSingleTrustModels}. Two combination approaches for the existing trust models are described and evaluated in Chapter~\ref{section:RobustnessOfCombinedTrustModels}. We conclude and recommend further work inspired by this research project in Chapter~\ref{section:Conclusion}.




\chapter{Related Work}
\label{section:Relatedwork}
\ifpdf
    \graphicspath{{Chapter1/Chapter1Figs/PNG/}{Chapter1/Chapter1Figs/PDF/}{Chapter1/Chapter1Figs/}}
\else
    \graphicspath{{Chapter1/Chapter1Figs/EPS/}{Chapter1/Chapter1Figs/}}
\fi

\section{Cheating Behaviors---The Attack Strategies}
\markboth{\MakeUppercase{\thechapter. Related Work }}{\thechapter. Related Work}
Various trust models have been proposed in different domains, such as P2P file sharing systems, ad-hoc networks, e-commerce \emph{etc.}.
G{\'o}mez M{\'a}rmol and Mart{\'\i}nez P{\'e}rez identified several common vulnerabilities of these trust models and provided recommendations for improving them (\cite{gomez2010towards}). Marmol and P{\'e}rez discussed several common attack strategies to the trust models for distributed systems (\cite{marmol2009security}). However, these studies did not evaluate attack strategies to trust models which are suitable for e-commerce.

In the context of e-commerce, cheating behaviors or attack strategies can be categories as \emph{Seller Cheating Behaviors} and \emph{Advisor Cheating Behaviors}.
Although the effectiveness of various attack strategies on trust models, including those suitable for e-commerce, has been studied in many other works (\emph{e.g.}, \cite{hussain2007state}, \cite{zhang2008detailed}, \cite{hoffman2009survey}, \cite{feng2010voting} and \cite{zhang2011extensive}), there is usually a lack of detailed experimental studies, especially the evaluation and comparison of a comprehensive set of unfair rating attack strategies on different trust models.

In the remaining section, the two types of cheating behaviors in the context of e-commerce will be discussed.

\subsection{Seller Cheating Behaviors}
In e-marketplaces, typical cheating behaviors from sellers, such as \emph{Reputation Lag}, \emph{Value Imbalance}, \emph{Re-entry}, \emph{Initial Window}, and \emph{Exit}, have been studied by Kerr and Cohen (\cite{kerr2006modeling}). A brief description of these seller cheating behaviors are given below (\cite{kerr09}).

\begin{itemize}
    \item \emph{Reputation Lag}: A common policy in many electronic marketplaces
is that the buyer pays before the seller ships the good. In this
scenario, a seller is likely to know that he intends to cheat from the
moment he receives payment. The buyer, however, will not know
for some time afterward, because of processing, shipping time, \emph{etc.}
Under some trust models, this presents an opportunity for a seller: he can
cheat a virtually unlimited number of times before his reputation is
updated to warn buyers of the new cheating activity.
    \item \emph{Value Imbalance}: In some trust models, all ratings are weighted
equally, regardless of the value of the transactions. This presents
an opportunity: a seller can honestly execute small sales, then use
the reputation gained to cheat on very large ones.
    \item \emph{Re-entry}: Users can create new accounts freely; in large markets, it is infeasible
to verify the identity of every trader. This presents the
opportunity for a dishonest trader to shed his bad reputation, starting
fresh by opening a new account. This is particularly dangerous
in systems that treat unknown sellers as preferable to disreputable
ones.
    \item \emph{Initial Window}: In some trust models, buyers rely only on their own
experience in evaluating sellers. Once a buyer has found trustworthy
sellers, this policy works well. Unfortunately, the buyer is vulnerable
until he finds those trustworthy sellers---he does not have
enough information to avoid cheaters.
    \item \emph{Exit}: If a seller cheats, it may damage his reputation, and hinder
his ability to engage in future sales. If the seller is planning to leave
the market, however, he has no further need for his good reputation.
Thus, he can cheat freely, to the maximum extent possible, without
consequence. This is an extremely difficult problem to combat, and
affects most trust models.
\end{itemize}

Kerr and Cohen assumed maximal cheating in their work of evaluating robustness of trust models against the above seller attacks: a cheating seller does not ship out his product thus no cost is incurred, and the buyer will learn the results only after the lag has lapsed (\cite{kerr09}).

Recent work by J$\o$sang and Golbeck identified more seller attack strategies and reduced all types of advisor cheating behaviors to Unfair Rating Attack (\cite{ajosang09}).
Particularly, Kerr and Cohen found combined seller attacks are able to defeat every investigated trust model (\cite{kerr09}). Researchers, especially those models' designers, might be tempted to argue that, cheating behaviors from sellers are possible to be handled by law and their models are still robust against advisors' unfair rating attack rather than sellers' attack strategies.

\subsection{Advisor Cheating Behaviors---Unfair Rating Attacks}
In this report, we argue that even though cheating behaviors from sellers are possible to be sanctioned by law, advisors' cheating behaviors are still able to defeat the existing trust models; thus, improving the robustness of the existing trust models for handling unfair ratings is urgently demanded.

To begin with, online transactions are essentially contracts: sellers are obliged to deliver products as specified by themselves and buyers are obliged to pay the specified amount of money. Therefore, most sellers' cheating behaviors can be considered as unlegal: in the real life, it is very common that buyers may sue their sellers if the delivered products are not as good as specified by the sellers according to the contract law.

Although sellers' cheating behaviors can be sanctioned by law, advisors' unfair ratings can only be considered as unethical rather than unlegal (\cite{ajosang09}), therefore there is an urgent demand to address the unfair rating problem. Our paper focuses on advisor cheating behaviors and below are a list of typical unfair rating attacks that may threaten the existing trust models in e-marketplaces. Note that some attack names are used interchangeably in both seller attacks and advisors' unfair rating attacks (\emph{e.g.}, Sybil Attack), in this report we refer to the latter.

\subsubsection{Constant Attack}
The simplest strategy from dishonest advisors is, constantly providing unfairly positive ratings to dishonest sellers while providing unfairly negative ratings to honest sellers. This simple attack is a baseline to test the basic effectiveness of different trust models in dealing with unfair ratings.

\subsubsection{Camouflage Attack}
Dishonest advisors may camouflage themselves as honest ones by providing fair ratings strategically. For example, advisors may provide fair ratings to build up their trustworthiness (according to certain trust models) at the early stage before providing unfair ratings. Intuitively, if trust models assume attackers' behaviors are constant and stable, they may be vulnerable to this type of attack.

\subsubsection{Whitewashing Attack}
In e-marketplaces, it is hard to establish buyers' identities: users can freely create a new account as a buyer. This presents an opportunity for a dishonest buyer to \emph{whitewash} his low trustworthiness (according to certain trust models) by starting a new account with the default initial trustworthiness value (0.5 in our investigated trust models).

\subsubsection{Sybil Attack}
When evaluating the robustness of trust models, it is usually assumed that the majority of buyers are honest. In our experiments, the aforementioned three types of attackers are minority compared with the remaining honest buyers. 

However, it is possible that dishonest buyers (unfair rating attackers) may form the majority of all the buyers in e-marketplaces. In this report, we use the term \emph{Sybil Attack}, which was initially proposed by Douceur, to describe the scenario where dishonest buyers have obtained larger amount of resources (buyer accounts) than honest buyers to constantly provide unfair ratings to sellers (\cite{douceur02}).

This attack can be considered as, dishonest buyers are more than honest buyers and they perform Constant Attack together.

\subsubsection{Sybil Camouflage Attack}
As the name suggests, this attack combines both Camouflage Attack and Sybil Attack: dishonest buyers are more than honest buyers and they perform Camouflage Attack together.

\subsubsection{Sybil Whitewashing Attack}
Similar to Sybil Camouflage Attack, this attack combines both Whitewashing Attack and Sybil Attack:
dishonest buyers are more than honest buyers and they perform Whitewashing Attack together. Intuitively, this new combined attack may pose a greater threat to trust models due to the presence of a larger number of attacker identities.

\subsubsection{Non-Sybil-based and Sybil-based Attack}
Obviously, under the Constant Attack, Camouflage Attack and Whitewashing Attack, the number of dishonest buyers is less than half of all the buyers in the market (minority). We refer to them as the \emph{Non-Sybil-based Attack}. On the contrary, the number of Sybil Attackers, Sybil Camouflage Attackers, or Sybil Whitewashing Attackers is greater than half of all the buyers (majority), and these attacks are referred to as the \emph{Sybil-based Attack}.

\section{Trust Models for Handling Unfair Rating---The Defense Mechanisms}
Various trust models have been proposed to deal with different unfair rating attacks. In the interest of fairness, we selected four representative models proposed during the year 2002---2011 that self-identified as applicable to e-marketplaces and robust against unfair rating attacks. In this chapter, we also classify them into two broad categories: \emph{Filtering-based} and \emph{Discounting-based}.

\subsection{Beta Reputation System (BRS)}
The Beta Reputation system (BRS) was proposed by J$\o$sang and Ismail to predict a seller's behavior in the next transaction based on the number of honest and dishonest transactions (the two events in the beta distribution: $[p, n]$, where $p$ and $n$ denote the number of received positive and negative ratings) he has conducted in the past (\cite{ajosang02}).

Whitby \emph{et al.} further proposed an iterative approach to filter out unfair ratings based on the \emph{majority rule} (\cite{whitby04}). According to this approach, if the calculated reputation of a seller based on the set of honest buyers (initially all buyers) falls in the rejection area ($q$ quantile or $1-q$ quantile) of the beta distribution of a buyer's ratings to that seller, this buyer will be filtered out from the set of honest buyers and all his ratings will be considered as unfair ratings since his opinions (ratings) are not consistent with the majority of the other buyers' opinions (the majority rule). Then the seller's reputation will be re-calculated based on the updated set of honest buyers, and the filtering process continues until the set of honest buyers eventually remains unchanged.

Obviously, the majority rule renders BRS vulnerable to Sybil-based Attack because the majority of buyers are dishonest and the other honest buyers' (the minority) ratings will be filtered out.

\subsection{iCLUB}
iCLUB is a recently proposed trust model by Liu \emph{et al.} in handling multi-nominal ratings (\cite{liu11}). It adopts the clustering approach and considers buyers' local and global knowledge about sellers to filter out unfair ratings.

For local knowledge, the buyer compares his ratings with advisors' ratings (normalized rating vectors) towards the \emph{target seller} (the seller under evaluation) by clustering. If an advisor's ratings are not in the cluster containing the buyer's ratings, they will be considered as not consistent with the buyer's opinions, and will be filtered out as unfair ratings. Obviously, comparing advisors' ratings with the buyer's own opinions is reliable since the buyer never lies to himself.

If transactions between the buyer and the target seller are too few (few evidence), the buyer will not be confident to rely on his local knowledge, and global knowledge will be used. The buyer will compare his and the advisors' ratings towards all the sellers excluding the target seller by performing clustering. A set of advisors who always have similar ratings with the buyer (in the same cluster) towards every seller are identified. Eventually, these advisors are used to filtered out the other untrustworthy advisors' ratings when evaluating all advisors' ratings to the target seller.

In general, buyers' local knowledge is more reliable than his global knowledge. This is because when the set of advisors whose opinions are always similar to the buyer's cannot be found, the global knowledge will use the majority rule to filter out unfair ratings; this may be vulnerable to Sybil-based Attack.

\subsection{Filtering-based Trust Models}
BRS and iCLUB filter out unfair ratings before aggregating the remaining fair ratings in evaluating a seller's reputation, therefore, we classified them as {\bf Filtering-based}. The reputation of the seller $S$, $\Gamma(S)$, is calculated as:

\begin{equation}
\label{eqn:filter}
 \Gamma(S) = \frac{\sum p_i + 1}{\sum p_i + \sum n_i + 2}
\end{equation}

where $p_i$ and $n_i$ are the number of positive and negative ratings from each advisor $i$ to the seller $S$ after unfair ratings are filtered out. When $S$ does not receive any ratings, his initial reputation is 0.5.

\subsection{TRAVOS}
Teacy \emph{et al.} proposed TRAVOS to evaluate the trustworthiness of advisors, $\tau_i$, and use $\tau_i$ to discount their ratings before aggregating these ratings to evaluate the target seller's reputation (\cite{teacy06}).

To evaluate an advisor's trustworthiness, first, a set of reference sellers are identified if these sellers' reputation are similar to the target seller's reputation as calculated by using this advisor's ratings towards them. Then the buyer will use the cumulative distribution function of beta distribution based on the total number of his positive and negative ratings to each reference seller to compute the trustworthiness of that advisor.

Compared with BRS, TRAVOS incorporates a buyer's personal transaction experiences with the target seller in the process of evaluating his advisors' trustworthiness. However, TRAVOS assumes the advisors' behaviors are constant; thus, this model may be vulnerable if the attackers camouflage themselves by giving fair ratings strategically before providing unfair ratings.

\subsection{Personalized}
Zhang and Cohen proposed a personalized approach to evaluate an advisor's trustworthiness $\tau_i$ in two aspects: private and public trust (\cite{jzhang06}).

To evaluate the private trust of an advisor, the buyer compares his ratings with the advisor's ratings to their commonly rated sellers. Greater disparity in the comparison indicates discounting of the advisor's trustworthiness to a larger extent.

Similarly, the public trust of an advisor is estimated by comparing the advisor's ratings with the majority of the other advisors' ratings towards their commonly rated sellers. Obviously, public trust adopts the majority rule in evaluating an advisor's trustworthiness and therefore may be vulnerable to Sybil-based Attack.

Since private trust is more reliable, when aggregating both private and public trust of an advisor, this model will allocate higher weightage to private trust if the buyer has more commonly rated sellers with the advisor (more evidence). When the number of such commonly rated sellers exceeds a certain threshold value (enough evidence), the buyer will only use the private trust to evaluate the advisor's trustworthiness more accurately.

\subsection{Discounting-based Trust Models} TRAVOS and Personalized calculate advisors' trustworthiness and use their trustworthiness to discount their ratings before aggregating them to evaluate a seller's reputation. Thus, we classified them as {\bf Discounting-based}. The reputation of the seller $S$, $\Gamma(S)$, is calculated as:

\begin{equation}
\label{eqn:discount}
 \Gamma(S) = \frac{\sum \tau_i \times p_i + 1}{\sum \tau_i \times p_i + \sum \tau_i \times n_i + 2}
\end{equation}

where $p_i$ and $n_i$ are the number of positive and negative ratings from each advisor $i$ to the seller $S$, and $\tau_i$ is the trustworthiness of the advisor $i$. When $S$ does not receive any ratings, his initial reputation is 0.5.



\chapter{Evaluation Method}
\label{section:Evaluationmethod}
\ifpdf
    \graphicspath{{Chapter2/Chapter2Figs/PNG/}{Chapter2/Chapter2Figs/PDF/}{Chapter2/Chapter2Figs/}}
\else
    \graphicspath{{Chapter2/Chapter2Figs/EPS/}{Chapter2/Chapter2Figs/}}
\fi

\section{The E-marketplace Testbed}
\label{subsection:Testbed}
\markboth{\MakeUppercase{\thechapter. Evaluation Method}}{\thechapter. Evaluation Method}

\subsection{The ART Testbed}
Our experiments were performed by simulating the transaction activities in the e-marketplace.
An existing testbed, ART, has been developed within the trust and
reputation community for both competition and experimentation (\cite{kFullam05}).

\begin{figure}[h]
\centering
\includegraphics[scale=1]{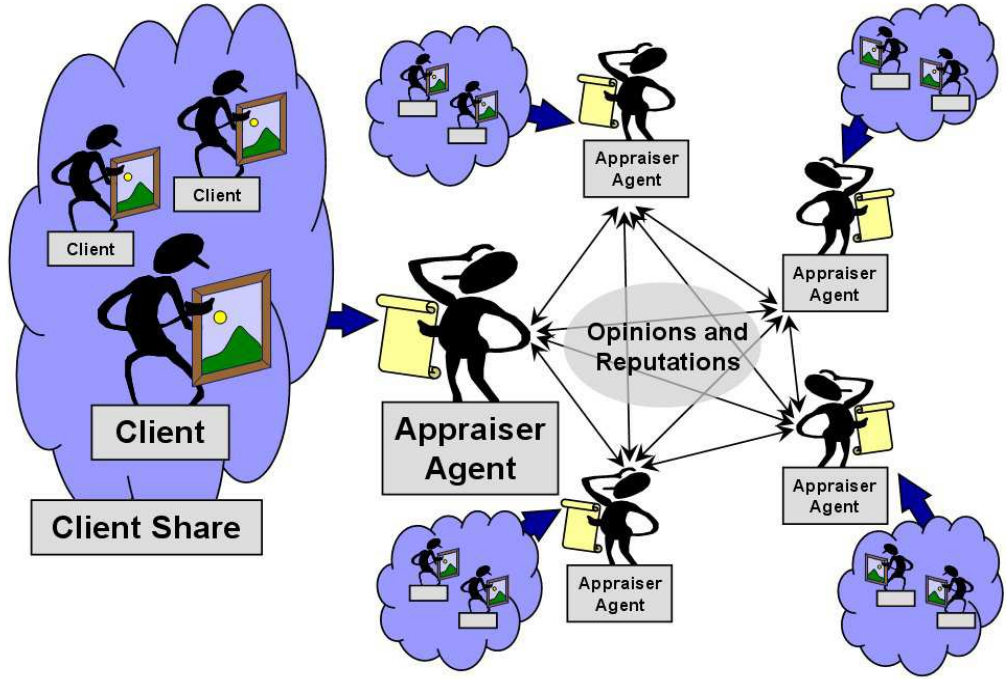}
\caption{Game overview of ART, demonstrating interactions between clients and appraiser agents (\cite{fullam2007agent}).}
\label{fig:art}
\end{figure}

The ART Testbed compares different trusting strategies as they act in combination.
In the art appraisal domain (Fig.~\ref{fig:art}), agents function as painting appraisers
with varying levels of expertise in different artistic eras. Clients request appraisals
for paintings from different eras; if an appraising agent does not have the expertise
to complete the appraisal, it can request opinions from other appraiser agents. Appraisers
receive more clients, and thus more profit, for producing more accurate
appraisals (\cite{fullam2007agent}).

While ART has much value, as mentioned in Chapter~\ref{section:introduction}, it is not suitable for carrying out experiments to compare robustness of trust models under different unfair rating attacks. For example, the role of agents as both buyers
and sellers makes it difficult to isolate the effects of individual buyer/seller strategies (\cite{kerr09}).

\subsection{The TREET Testbed}
Kerr and Cohen proposed TREET, an experimentation
and evaluation testbed based directly on that used in their investigations into security
vulnerabilities in trust and reputation systems for e-marketplaces (\cite{kerr2009experimental} and \cite{kerr2010treet}).

\begin{figure}[h]
\centering
\includegraphics[scale=0.65]{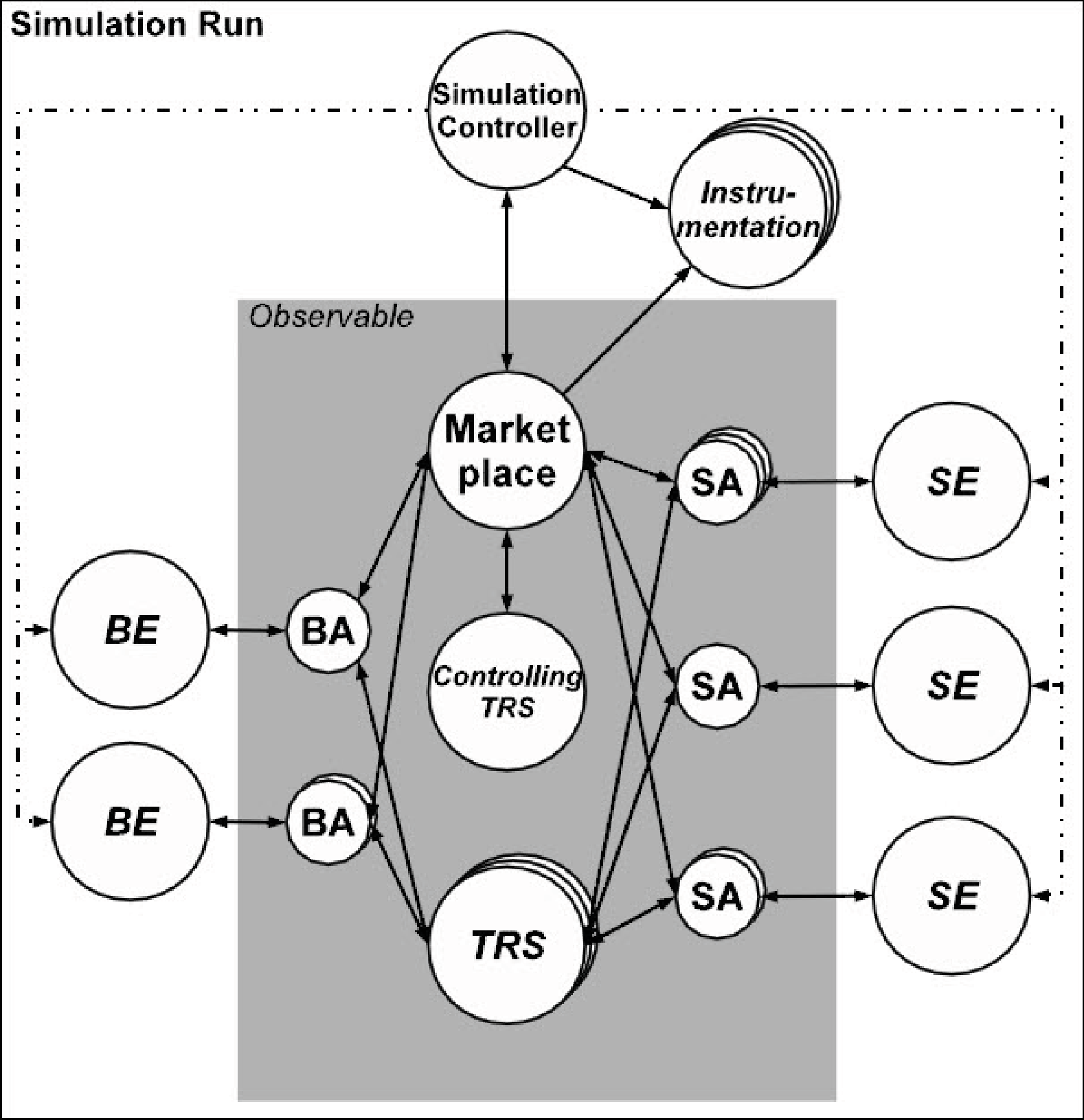}
\caption{The TREET architecture (\cite{kerr2010treet}).}
\label{fig:treet}
\end{figure}

The architecture is depicted in Fig.~\ref{fig:treet}. In this diagram, BA and
SA refer to Buying Account and Selling Account respectively. BE and SE refer to
Buying Entity and Selling Entity respectively. All components labelled in italic text
are components that are intended to be provided/modified by investigators making
use of the testbed. The gray box denotes those components that are observable by
marketplace participants, although this does not imply complete visibility.
Each complete run of the testbed is represented by a SimulationRun, into which
the necessary arguments and objects are passed. A SimulationRun is responsible for
setup and configuration of a run---creation of the product set, initialization of components,
\emph{etc.}---and initiating the Simulation Controller. A set of numerous tests can
be executed by creating multiple instances of SimulationRun.
A Simulation Controller is responsible for actual execution of a simulation run.
The controller triggers each of the day's events in turn, signalling the appropriate
parties when they are required to take action.
The scenario makes use of a single centralized marketplace, represented by a Marketplace object.
All offers, acceptances, and payments are made through the Marketplace. All accounts reside in the
Marketplace, and requests to open accounts are processed through it. (\cite{kerr2010treet})

Although TREET is suitable for evaluation of seller attack strategies, it is not flexible in incorporating various unfair rating attacks.

\subsection{The ``Duopoly Market" Testbed}
In light of the limitations of ART and TREET, we designed and developed an e-marketplace testbed, which is extendable via incorporating new trust models for handling unfair ratings or new advisor attack strategies (unfair rating attacks).

\begin{figure}[t]
\centering
\includegraphics[scale=0.78]{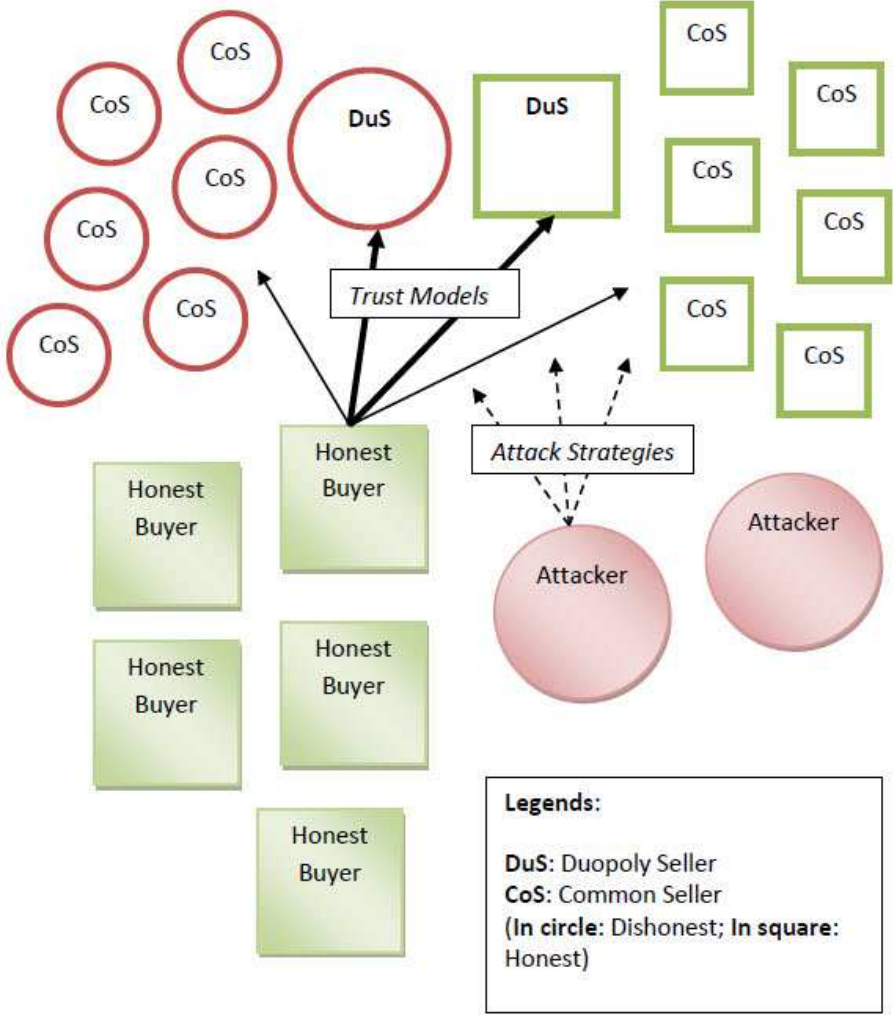}
\caption{The structure of the ``Duopoly Market" Testbed.}
\label{fig:duotestbed}
\end{figure}

In our e-marketplace testbed, there are 10 dishonest sellers and 10 honest sellers. To make the comparison more obvious, we considered a ``Duopoly Market": there are two sellers in the market that take up a large portion of the total transaction volume in the market. We assumed a reasonable competition scenario: one duopoly seller (\emph{dishonest duopoly seller}) tries to beat his competitor (\emph{honest duopoly seller}) in the transaction volume by hiring or collaborating with dishonest buyers to perform unfair rating attacks. We refer to the remaining sellers (excluding the duopoly sellers) as \emph{common sellers}.

Typically, trust models are most effective when 30\% of buyers are dishonest (\cite{whitby04}). To ensure the best case for the trust models, we added 6 dishonest buyers (attackers) and 14 honest buyers in the market for Non-Sybil-based Attack, and switch their values for Sybil-based Attack.

The structure of the ``Duopoly Market" Testbed is shown in Fig.~\ref{fig:duotestbed}.
The entire simulation will last for 100 days. On each day, each buyer chooses to transact with one seller once.
Since most trust models are more effective when every advisor has transaction experiences with many different sellers, we assumed that there is a probability of 0.5 that buyers will transact with the duopoly sellers while there is another probability of 0.5 that buyers will transact with each common seller randomly. The value of 0.5 also implies that the duopoly sellers take up half of all the transactions in the market. When deciding on which duopoly seller to transact with, honest buyers use trust models to calculate their reputation values and transact with the one with the higher value, while dishonest buyers choose sellers according to their attacking strategies. After each transaction, honest buyers provide fair ratings, whereas dishonest buyers provide ratings according to their attack strategies.

The key parameters with their values in the e-marketplace testbed are summarized as follows:
\begin{itemize}
    \item \emph{Number of honest duopoly seller}: 1
    \item \emph{Number of dishonest duopoly seller}: 1
    \item \emph{Number of honest common seller}: 9
    \item \emph{Number of dishonest common seller}: 9
    \item \emph{Number of honest buyer/advisor ($|B^{H}|$)}: 14 (Non-Sybil-based Attack) or 6 (Sybil-based Attack)
    \item \emph{Number of dishonest buyer/advisor or attacker ($|B^{D}|$)}: 6 (Non-Sybil-based Attack) or 14 (Sybil-based Attack)
    \item \emph{Number of simulation days ($L$)}: 100
    \item \emph{The ratio of duopoly sellers' transactions to all transactions ($r$)}: 0.5
\end{itemize}

\section{The Trust Model Robustness Metric}
\markboth{\MakeUppercase{\thechapter. Evaluation Method}}{\thechapter. Evaluation Method}

To evaluate the robustness of different trust models, we compared the transaction volumes of the duopoly sellers. Obviously, the more robust the trust model, the larger the transaction volume difference between the honest and dishonest duopoly seller. The robustness of a trust model (defense, $Def$) against an attack model ($Atk$) is defined as:

\begin{equation}
\label{eqn:robustness}
\Re(Def, Atk)= \frac{|Tran(S^{H})|-|Tran(S^{D})|}{|B^{H}|\times L \times r}
\end{equation}

where
$|Tran(S^{H})|$ and $|Tran(S^{D})|$ denote the total transaction volume of the honest and dishonest duopoly seller,
and the values of key parameters in the e-marketplace testbed $|B^{H}|$, $L$, and $r$ are given in Chapter~\ref{subsection:Testbed}.

If a trust model $Def$ is \emph{completely robust} against a certain attack $Atk$, theoretically $\Re(Def, Atk) = 1$. It means the reputation of the honest duopoly seller is always higher than that of the dishonest duopoly seller as calculated by the trust model, so honest buyers will always transact with the honest duopoly seller.
On the contrary, $\Re(Def, Atk) = -1$ indicates, the trust model always suggests honest buyers to transact with the dishonest duopoly seller, and $Def$ is \emph{completely vulnerable} to $Atk$.

When $\Re(Def, Atk) > 0$, the greater the value is, the more robust $Def$ is against $Atk$. When $\Re(Def, Atk) < 0$, the greater the absolute value is, the more vulnerable $Def$ is to $Atk$. It should be noted that, when $Def$ is completely robust against or vulnerable to $Atk$, in our experiments $\Re(Def, Atk)$ can be slightly around 1 or -1 because the probability to transact with the duopoly sellers may not be exactly 0.5 in the actual simulation process.

In Eq.~\ref{eqn:robustness}, the denominator denotes the transaction volume difference between the honest and dishonest duopoly seller when the trust model ($Def$) is completely robust against or vulnerable to a certain attack ($Atk$): all the honest buyers ($B^{H}$) always transact with the duopoly honest seller ($S^{H}$, when completely robust) or duopoly dishonest seller ($S^{D}$, when completely vulnerable) in the 100 days with a probability of 0.5 to transact with the duopoly sellers.

In our experiments, the denominator is 700 ($14\times100\times0.5$) if $Atk$ is Non-Sybil-based Attack, or 300 ($6\times100\times0.5$) if $Atk$ is Sybil-based Attack.



\chapter{Robustness of Single Trust Models}
\label{section:RobustnessOfSingleTrustModels}
\ifpdf
    \graphicspath{{Chapter3/Chapter3Figs/PNG/}{Chapter3/Chapter3Figs/PDF/}{Chapter3/Chapter3Figs/}}
\else
    \graphicspath{{Chapter3/Chapter3Figs/EPS/}{Chapter3/Chapter3Figs/}}
\fi

\section{Experiments}
\markboth{\MakeUppercase{\thechapter. Robustness of Single Trust Models }}{\thechapter. Robustness of Single Trust Models}
This chapter evaluates the robustness of all the trust models against all the attack strategies covered in Chapter~\ref{section:Relatedwork} with the e-marketplace testbed described in Chapter~\ref{section:Evaluationmethod}. In our experiments, when models require parameters we have used values provided by the authors in their own works wherever possible.

The experiments were performed 50 times, and the mean and standard deviation of the 50 results are shown in Table~\ref{table:RobustnessSingle} in the form of ($mean \pm std$). The robustness of all the single trust models against each attack is discussed in the remaining of this chapter.

\begin{table}[t]
\caption{Robustness of single trust models against attacks. Every entry denotes the mean and standard deviation of the robustness values of trust model against attack}
\vspace{5 mm}
\label{table:RobustnessSingle}
\centering
\scriptsize
\begin{tabular}{|l|c|c|c|c|c|c|}
\hline &  Constant & Camouflage & Whitewashing & Sybil & Sybil Cam & Sybil WW \\
\hline
BRS                    &  0.84$\pm$0.03	 &  0.87$\pm$0.04	&  -0.48$\pm$0.08	    &  -0.98$\pm$0.09	    &  -0.63$\pm$0.08     &  -0.60$\pm$0.10	  \\
iCLUB	               &  1.00$\pm$0.04	 &  0.98$\pm$0.03	&  $~$0.81$\pm$0.10	    &  -0.09$\pm$0.33	    &  $~$0.95$\pm$0.11   &  -0.16$\pm$0.26	  \\
TRAVOS                 &  0.96$\pm$0.04	 &  0.88$\pm$0.04	&  $~$0.98$\pm$0.04	    &  $~$0.66$\pm$0.10	    &  -0.60$\pm$0.09     &  -1.00$\pm$0.08	  \\
Personalized           &  0.99$\pm$0.04	 &  1.01$\pm$0.03	&  $~$0.99$\pm$0.04	    &  $~$0.84$\pm$0.12	    &  $~$0.67$\pm$0.09   &  -1.00$\pm$0.11	  \\
\hline
\multicolumn{7}{l}{*Sybil Cam: Sybil Camouflage Attack; Sybil WW: Sybil Whitewashing Attack}\\
\end{tabular}
\end{table}

\section{Robustness to Constant Attack}
It is observed that all the trust models are robust against this baseline attack.

\begin{figure}[t!]
\centering
\includegraphics[scale=0.7]{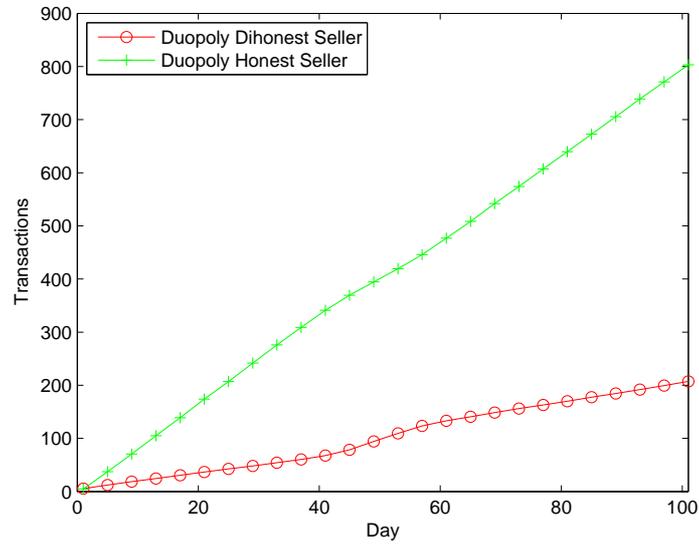}
\caption{BRS vs. Constant Attack}
\label{fig:brs_constant}
\end{figure}

\begin{figure}[t!]
\centering
\includegraphics[scale=0.7]{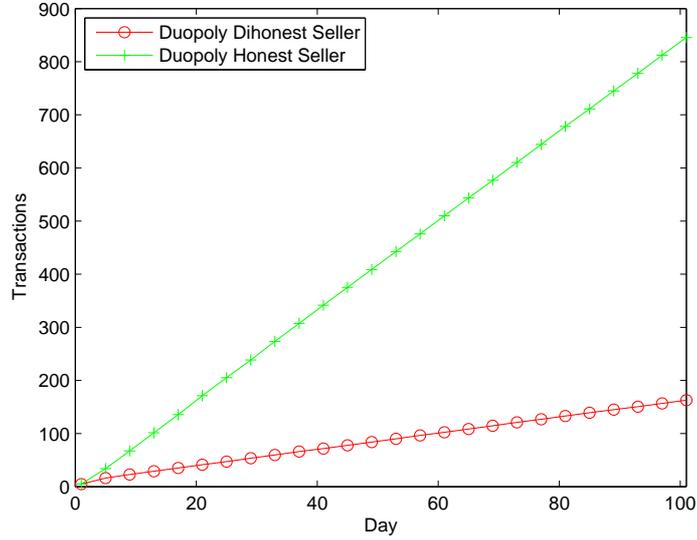}
\caption{iCLUB vs. Constant Attack}
\label{fig:iclub_constant}
\end{figure}

\begin{figure}[t!]
\centering
\includegraphics[scale=0.7]{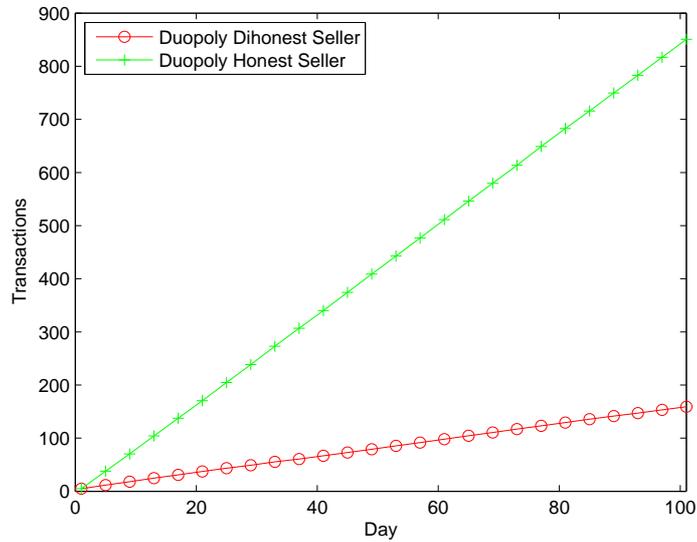}
\caption{TRAVOS vs. Constant Attack}
\label{fig:travos_constant}
\end{figure}

\begin{figure}[t!]
\centering
\includegraphics[scale=0.7]{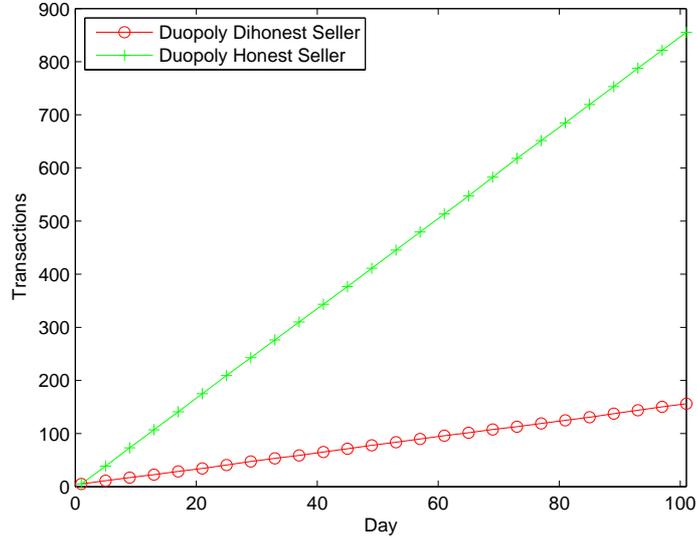}
\caption{Personalized vs. Constant Attack}
\label{fig:personalized_constant}
\end{figure}

Consistent with Whitby \emph{et al.}'s experimental results, our experiments also showed BRS is not completely robust against Constant Attack (\cite{whitby04}). Fig.~\ref{fig:brs_constant}---Fig.~\ref{fig:personalized_constant} depict under Constant Attack, how the transactions of the duopoly sellers grow day after day when BRS, iCLUB, TRAVOS and Personalized are used by honest buyers to decide which duopoly seller to transact with. The transaction volume difference between the honest and dishonest duopoly seller on Day 100 (around 700) indicates that iCLUB, TRAVOS and Personalized are completely robust against Constant Attack.

Space prevents the inclusion of such figures for every trust model; throughout this paper, all key data are presented in Table~\ref{table:RobustnessSingle} and Table~\ref{table:RobustnessCombine} and we use charts where illustration is informative.

\section{Robustness to Camouflage Attack}
In this experiment, Camouflage Attackers give fair ratings to all the common sellers to establish their trustworthiness before giving unfair ratings to all sellers (with a probability of 0.5 to transact with the duopoly sellers).

From the results of Table~\ref{table:RobustnessSingle}, without enough attackers, Camouflage Attack does not threaten the trust models very much.

\section{Robustness to Whitewashing Attack}
In our experiment, each Whitewashing Attacker provides one unfair rating on one day and starts with a new buyer account on the next day.

\begin{figure}[t!]
\centering
\includegraphics[scale = 0.7]{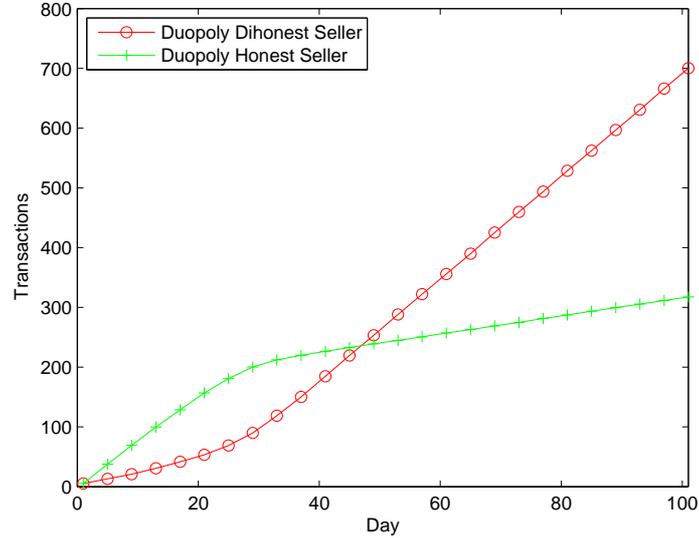}
\caption{BRS vs. Whitewashing Attack} \label{fig:brs_whitewashing}
\end{figure}

The value $\Re(BRS, Whitewashing) = -0.48$ in Table~\ref{table:RobustnessSingle} shows BRS is vulnerable to this attack. According to Fig.~\ref{fig:brs_whitewashing}, the honest duopoly seller has more transactions than the dishonest one at the beginning. However, after some time (around Day 45) the dishonest duopoly seller's transaction volume exceeds his competitor. In fact, after some time the calculated reputation of a seller will more easily fall in the rejection area of the beta distribution of an honest buyer's single accumulated ratings (single $[p, 0]$ to an honest seller and single $[0, n]$ to a dishonest seller, where $p$ and $n$ become very large as transaction experiences accumulate) rather than Whitewashing Attackers' multiple one-transaction ratings (multiple $[0, 1]$ to an honest sellers and multiple $[1, 0]$ to a dishonest seller).

The other trust models are robust against Whitewashing Attack.

\section{Robustness to Sybil Attack}
\begin{figure}[t!]
\centering
\includegraphics[scale = 0.7]{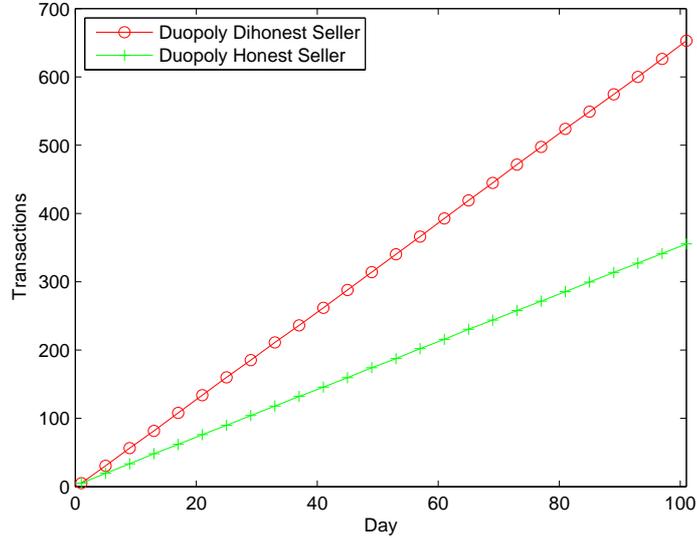}
\caption{BRS vs. Sybil Attack} \label{fig:brs_sybil}
\end{figure}

As described in Chapter~\ref{section:Relatedwork}, BRS is completely vulnerable to Sybil Attack due to its employed majority-rule (Fig.~\ref{fig:brs_sybil}).

The robustness of iCLUB is not stable as indicated by its standard deviation of 0.33. To explain, an honest buyer can rely on his local knowledge to always transact with one duopoly seller while using the global knowledge, which is wrong when majority of advisors are attackers, to evaluate the reputation of the other duopoly seller. The duopoly seller to always transact with can be either honest or dishonest as long as his reputation is always higher than that of his competitor, which is possible in either case.

Besides, TRAVOS and Personalized are not completely robust against Sybil Attack. This is due to the lack of transactions among different buyers and sellers at the beginning.

For TRAVOS, at the beginning it is hard to find common reference sellers for the buyer and the advisor so the discounting is not effective (we refer to this phenomenon as \emph{soft punishment}). When majority are dishonest buyers, their aggregated ratings will overweigh honest buyers' opinions.

For instance, if the trustworthiness of each dishonest and honest buyer are 0.4 and 0.6, and all buyers provides only one rating to a particular seller, according to Eq.~\ref{eqn:discount}, the reputation of an honest seller is $0.41 < 0.5$ ($0.41 = (0.6\times6+1)/(0.4\times14+0.6\times6+2)$) and that of a dishonest seller is $0.59 > 0.5$ ($0.59 = (0.4\times14+1)/(0.4\times14+0.6\times6+2)$); both suggest inaccurate decisions.
However, if a Discounting-based model is able to discount the trustworthiness of a dishonest buyer to a larger extent, say 0.1, while promote that of an honest buyer to a larger extent, say 0.9, the evaluation of sellers' reputation will become accurate.

For Personalized, at the beginning the buyer will more rely on public trust to evaluate the trustworthiness of an advisor, which is inaccurate when majority of buyers are dishonest.

\begin{figure}[t!]
\centering
\includegraphics[scale=0.7]{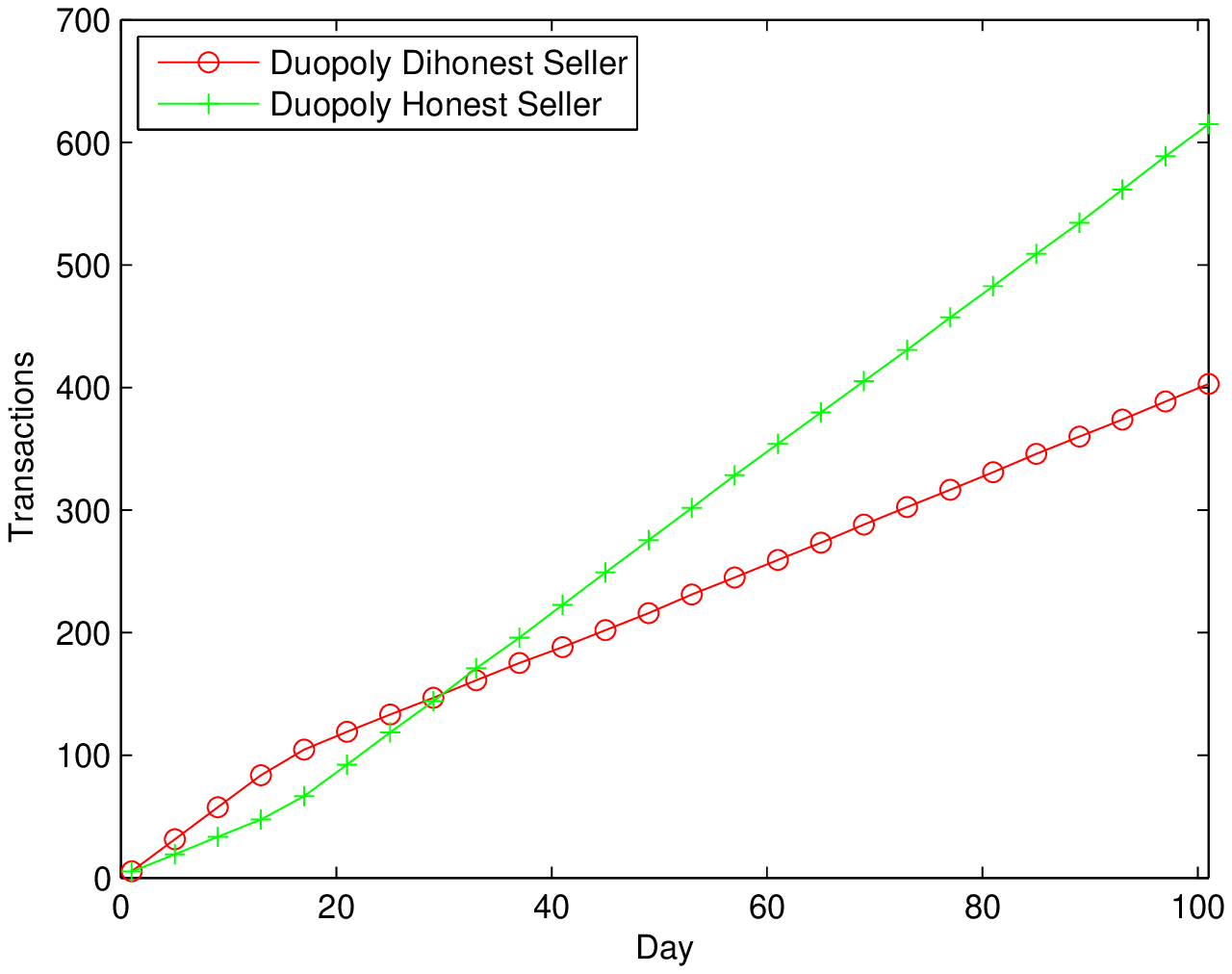}
\caption{TRAVOS vs. Sybil Attack}
\label{fig:travos_sybil}
\end{figure}

\begin{figure}[t!]
\centering
\includegraphics[scale=0.7]{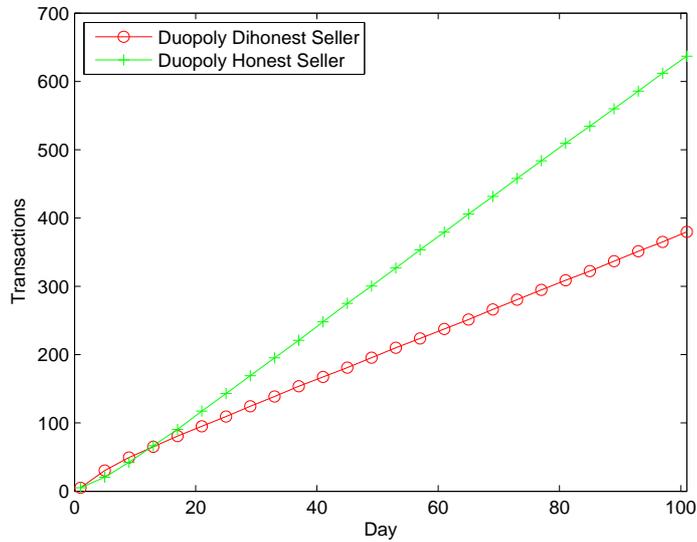}
\caption{Personalized vs. Sybil Attack}
\label{fig:personalized_sybil}
\end{figure}

Fig.~\ref{fig:travos_sybil} and Fig.~\ref{fig:personalized_sybil} show that, as transactions among different buyers and sellers grow, TRAVOS becomes more effective in discounting advisors' trustworthiness and Personalized tends to use private trust to accurately evaluate advisors' trustworthiness.

\section{Robustness to Sybil Camouflage Attack}
\begin{figure}[t!]
\centering
\includegraphics[scale=0.7]{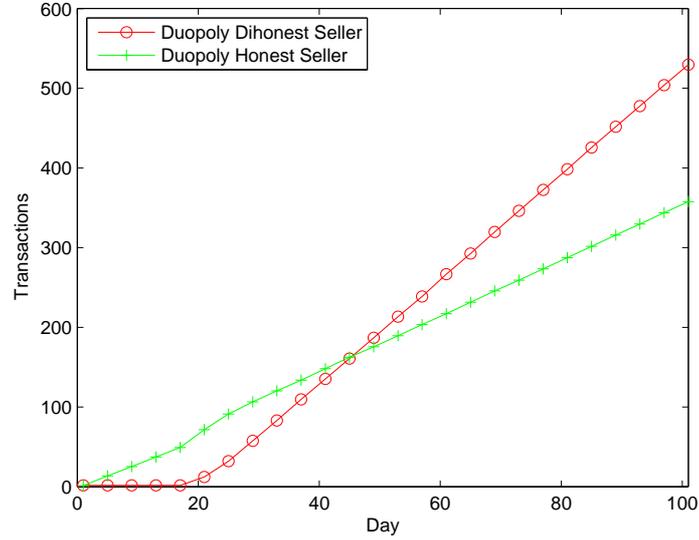}
\caption{BRS vs. Sybil Camouflage Attack}
\label{fig:brs_sybil_camouflage}
\end{figure}

Unlike Sybil Attack, Sybil Camouflage Attack is unable to render BRS completely vulnerable. Based on Fig.~\ref{fig:brs_sybil_camouflage}, this is because at the beginning attackers camouflage themselves as honest ones by providing fair ratings, where BRS is always effective. After attackers stop camouflaging, the duopoly dishonest seller's transaction volume will soon exceed his competitor.

\begin{figure}[t!]
\centering
\includegraphics[scale=0.7]{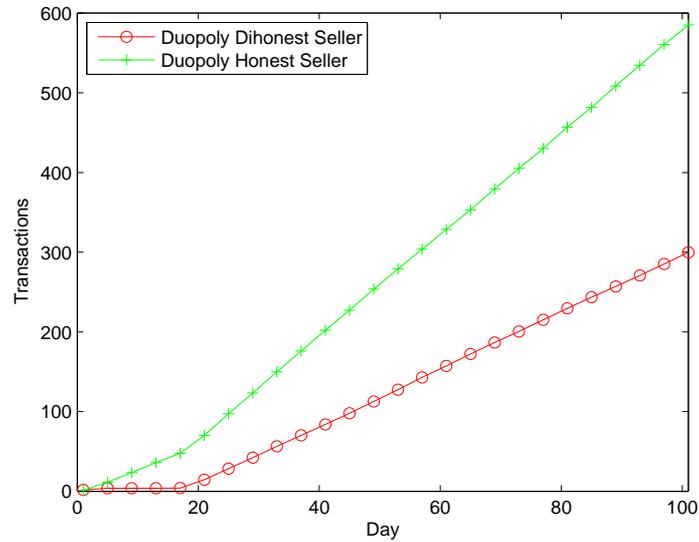}
\caption{iCLUB vs. Sybil Camouflage Attack}
\label{fig:iclub_sybil_camouflage}
\end{figure}

\begin{figure}[t!]
\centering
\includegraphics[scale=0.7]{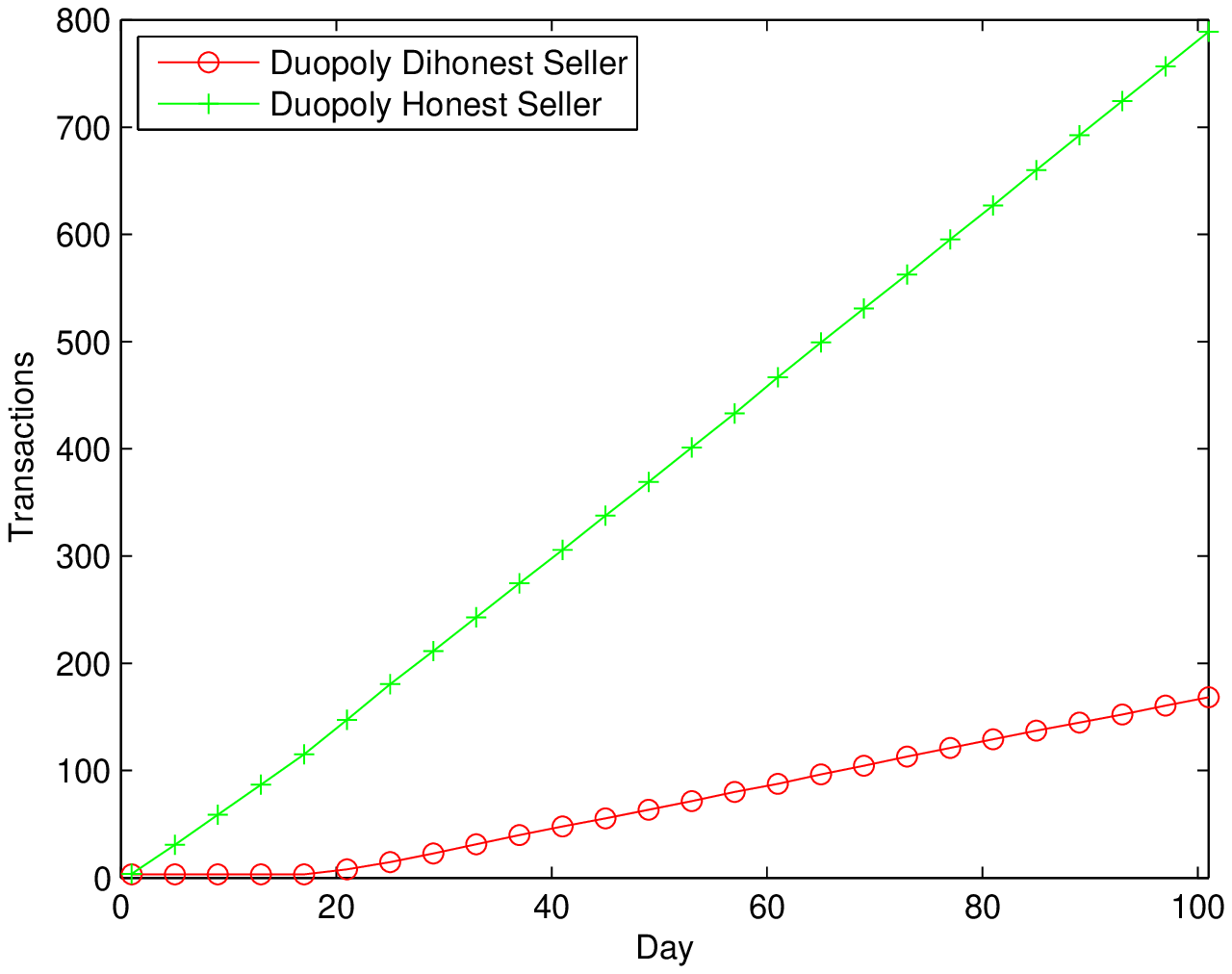}
\caption{TRAVOS vs. Camouflage Attack}
\label{fig:travos_camouflage}
\end{figure}

\begin{figure}[t!]
\centering
\includegraphics[scale=0.7]{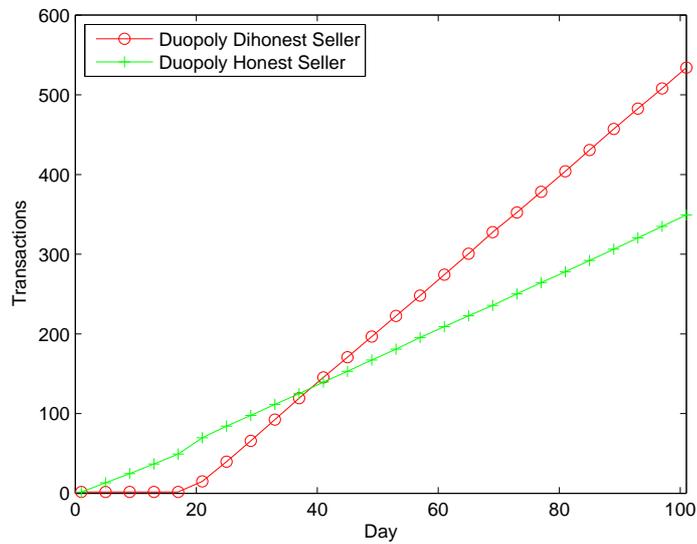}
\caption{TRAVOS vs. Sybil Camouflage Attack}
\label{fig:travos_sybil_camouflage}
\end{figure}

iCLUB is completely robust to Sybil Camouflage Attack. According to Fig.~\ref{fig:iclub_sybil_camouflage}, during the camouflaging stage, the honest duopoly seller will only transact with honest buyers. After attackers stop camouflaging, only the reliable local knowledge will be used by honest buyers to evaluate the trustworthiness of the honest duopoly seller (of high value), and honest buyers will continue to transact with him.

Compared with Camouflage and Sybil Attack, Personalized becomes less robust against Sybil Camouflage Attack. This is because the public and private trust of attackers have not been discounted to a large extent right after they complete the camouflaging stage (soft punishment). When the majority are attackers, their aggregated ratings will overweigh honest buyers' opinions. After attackers stop camouflaging, their private trust will continue to drop and Personalized will be effective.

Compared with Camouflage Attack, TRAVOS becomes vulnerable to Sybil Camouflage Attack: although TRAVOS will inaccurately promote the trustworthiness of a Camouflage Attacker (most are slightly larger than 0.5), when majority are honest buyers, the aggregated ratings from attackers are still not able to overweigh honest buyers' opinions. However, under Sybil Camouflage Attack, when majority are dishonest buyers, these attackers' aggregated ratings will easily overweigh honest buyers' opinions and render TRAVOS vulnerable.

Fig.~\ref{fig:travos_camouflage} and Fig.~\ref{fig:travos_sybil_camouflage} clearly show the difference of the robustness of TRAVOS against  Camouflage Attack and Sybil Camouflage Attack.

\section{Robustness to Sybil Whitewashing Attack}
This is the strongest attack: it can defeat every single trust model as observed from Table~\ref{table:RobustnessSingle}.

Similar to Sybil Attack, the robustness of iCLUB against Sybil Whitewashing Attack is still not stable.

Compared with Whitewashing Attack, BRS is still vulnerable to Sybil Whitewashing Attack while TRAVOS and Personalized change dramatically from completely robust to completely vulnerable.

\begin{figure}[t!]
\centering
\includegraphics[scale=0.7]{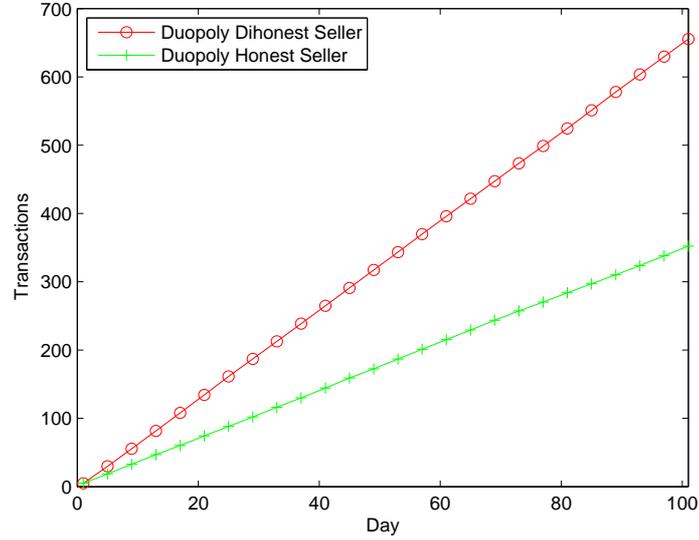}
\caption{TRAVOS vs. Sybil Whitewashing Attack}
\label{fig:travos_sybil_whitewashing}
\end{figure}

For TRAVOS, since every whitewashing attacker provides only one rating to a duopoly seller, buyer cannot find reference seller to effectively discount the trustworthiness of whitewashing attackers to a large extent. When majority are soft punished dishonest buyers, TRAVOS will always suggest honest buyers to transact with the dishonest duopoly seller. The complete vulnerability of TRAVOS to Sybil Whitewashing Attack is also depicted in Fig.~\ref{fig:travos_sybil_whitewashing}.

For Personalized, since every whitewashing attacker provides only one rating to a duopoly seller, the buyer cannot find enough commonly rated sellers and will heavily rely on public trust to evaluate the trustworthiness of an advisor, which is inaccurate when majority of buyers are dishonest. Therefore, similar to TRAVOS, the trustworthiness of whitewashing attacker cannot be discounted to a large extent and the soft punishment renders Personalized completely vulnerable.

It is also noted that although discounting-based TRAVOS and Personalized are robust against Whitewashing, Camouflage, and Sybil Attack, their robustness drops to different extents when facing Sybil Whitewashing and Sybil Camouflage Attack.

Based on our results demonstrated in Table~\ref{table:RobustnessSingle}, we conclude that, none of our investigated single trust models is robust against all the six attacks. Therefore, there is a demand to address the threats from all these attacks.



\chapter{Robustness of Combined Trust Models}
\label{section:RobustnessOfCombinedTrustModels}
\ifpdf
    \graphicspath{{Chapter4/Chapter4Figs/PNG/}{Chapter4/Chapter4Figs/PDF/}{Chapter4/Chapter4Figs/}}
\else
    \graphicspath{{Chapter4/Chapter4Figs/EPS/}{Chapter4/Chapter4Figs/}}
\fi

\section{Combining Trust Models}
\markboth{\MakeUppercase{\thechapter. Robustness of Combined Trust Models}}{\thechapter. Robustness of Combined Trust Models}
Based on the results of Table~\ref{table:RobustnessSingle}, Discounting-based trust models may change from vulnerable to robust if some attackers' ratings can be filtered out by Filtering-based models to reduce the effect of Sybil-based Attack to that of Non-Sybil-based Attack.

On the other hand, based on analysis in Chapter~\ref{section:RobustnessOfSingleTrustModels}, under most attacks Discounting-based models are still able to discount the trustworthiness of dishonest buyers to lower than 0.5 (although only slightly). Intuitively, filtering out ratings from advisors with lower trustworthiness may be a promising pre-filtering step before using Filtering-based models.

Therefore, we combine trust models from different categories to evaluate their new robustness to the same set of attacks. Generally, there are two approaches for combination: {\bf Filter-then-Discount} and {\bf Discount-then-Filter} (Fig.~\ref{combine}).
Details are given below.

\begin{figure}[h]
\centering
\includegraphics[scale = 0.45]{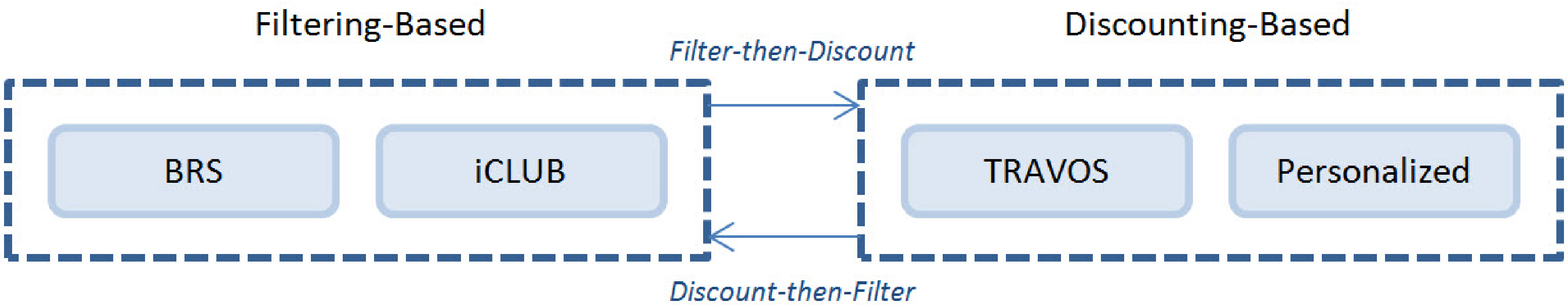}
\caption{Combining Trust Models} \label{combine}
\end{figure}

\vspace{5 mm}

\begin{table}[t]
\caption{Robustness of combined trust models against attacks. Every entry denotes the mean and standard deviation of the robustness values of trust model against attack}
\vspace{5 mm}
\label{table:RobustnessCombine}
\centering
\scriptsize
\begin{tabular}{|l|c|c|c|c|c|c|}
\hline &  Constant & Camouflage & Whitewashing & Sybil & Sybil Cam & Sybil WW \\
\hline
\multicolumn{7}{|c|}{Filter-then-Discount} \\
\hline
BRS + TRAVOS 	       &  0.89$\pm$0.06	 &  0.87$\pm$0.03	    &  -0.55$\pm$0.10	    &  -1.01$\pm$0.11	    &  -0.55$\pm$0.09    &  -0.59$\pm$0.11	    \\
BRS + Personalized     &  0.89$\pm$0.06	 &  0.88$\pm$0.03	    &  -0.34$\pm$0.05	    &  -0.96$\pm$0.07	    &  -0.53$\pm$0.08    &  -0.58$\pm$0.08	    \\
iCLUB + TRAVOS	       &  0.96$\pm$0.03	 &  0.98$\pm$0.04	    &  $~$0.95$\pm$0.04	    &  $~$0.85$\pm$0.08	    &  $~$0.97$\pm$0.10  &  $~$0.70$\pm$0.12	\\
iCLUB + Personalized   &  0.98$\pm$0.03	 &  0.99$\pm$0.03	    &  $~$0.92$\pm$0.06	    &  $~$0.88$\pm$0.13	    &  $~$0.98$\pm$0.09  &  $~$0.67$\pm$0.13	\\
\hline
\multicolumn{7}{|c|}{Discount-then-Filter} \\
\hline
TRAVOS + BRS	       &  0.95$\pm$0.03	 &  0.86$\pm$0.06	    &  $~$0.98$\pm$0.04	     &  $~$0.91$\pm$0.06	&  -0.57$\pm$0.12    &  $~$0.98$\pm$0.10	\\
TRAVOS + iCLUB	       &  0.95$\pm$0.04	 &  0.92$\pm$0.03	    &  $~$0.93$\pm$0.03	     &  $~$0.91$\pm$0.12	&  $~$0.91$\pm$0.10  &  $~$0.94$\pm$0.12	\\
Personalized + BRS     &  0.99$\pm$0.03	 &  0.98$\pm$0.03	    &  $~$1.01$\pm$0.03	     &  $~$0.96$\pm$0.11	&  $~$0.87$\pm$0.08  &  $~$1.00$\pm$0.10	\\
Personalized + iCLUB   &  0.97$\pm$0.04	 &  0.95$\pm$0.02	    &  $~$0.98$\pm$0.04	     &  $~$0.92$\pm$0.09	&  $~$0.94$\pm$0.09  &  $~$0.93$\pm$0.07	\\
\hline
\multicolumn{7}{l}{*Sybil Cam: Sybil Camouflage Attack; Sybil WW: Sybil Whitewashing Attack}\\
\end{tabular}
\end{table}

\subsection{Approach 1---Filter-then-Discount:}
\begin{enumerate}
  \item Use a Filtering-based trust model to filter out unfair ratings;
  \item Use a Discounting-based trust model to aggregate discounted ratings to calculate sellers' reputation.
\end{enumerate}

\subsection{Approach 2---Discount-then-Filter:}
\begin{enumerate}
  \item Use a Discounting-based trust model to calculate each advisor $i$'s trustworthiness $\tau_i$;
  \item If $\tau_i < \epsilon$, remove $i$'s all ratings ($\epsilon = 0.5$ in our experiment);
  \item Use a Filtering-based trust model to filter out unfair ratings before aggregating the remaining ratings to calculate sellers' reputation.
\end{enumerate}

\begin{figure}[!t]
\centering
\includegraphics[scale=0.7]{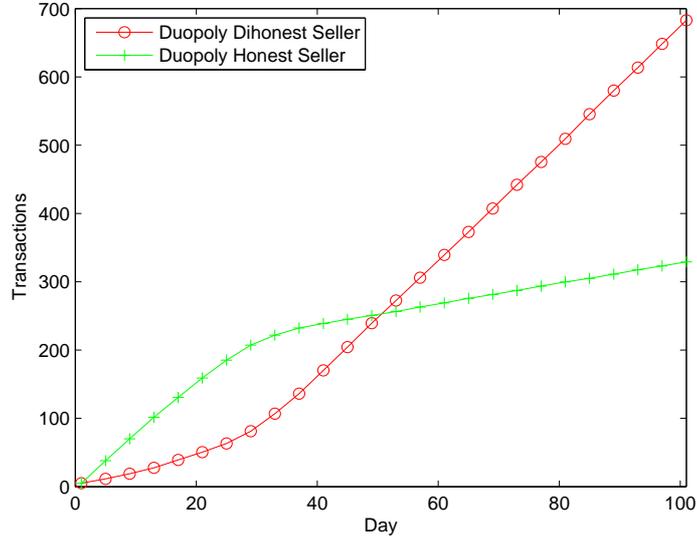}
\caption{BRS + TRAVOS vs. Whitewashing Attack}
\label{fig:bt_whitewashing}
\end{figure}

\begin{figure}[!t]
\centering
\includegraphics[scale=0.7]{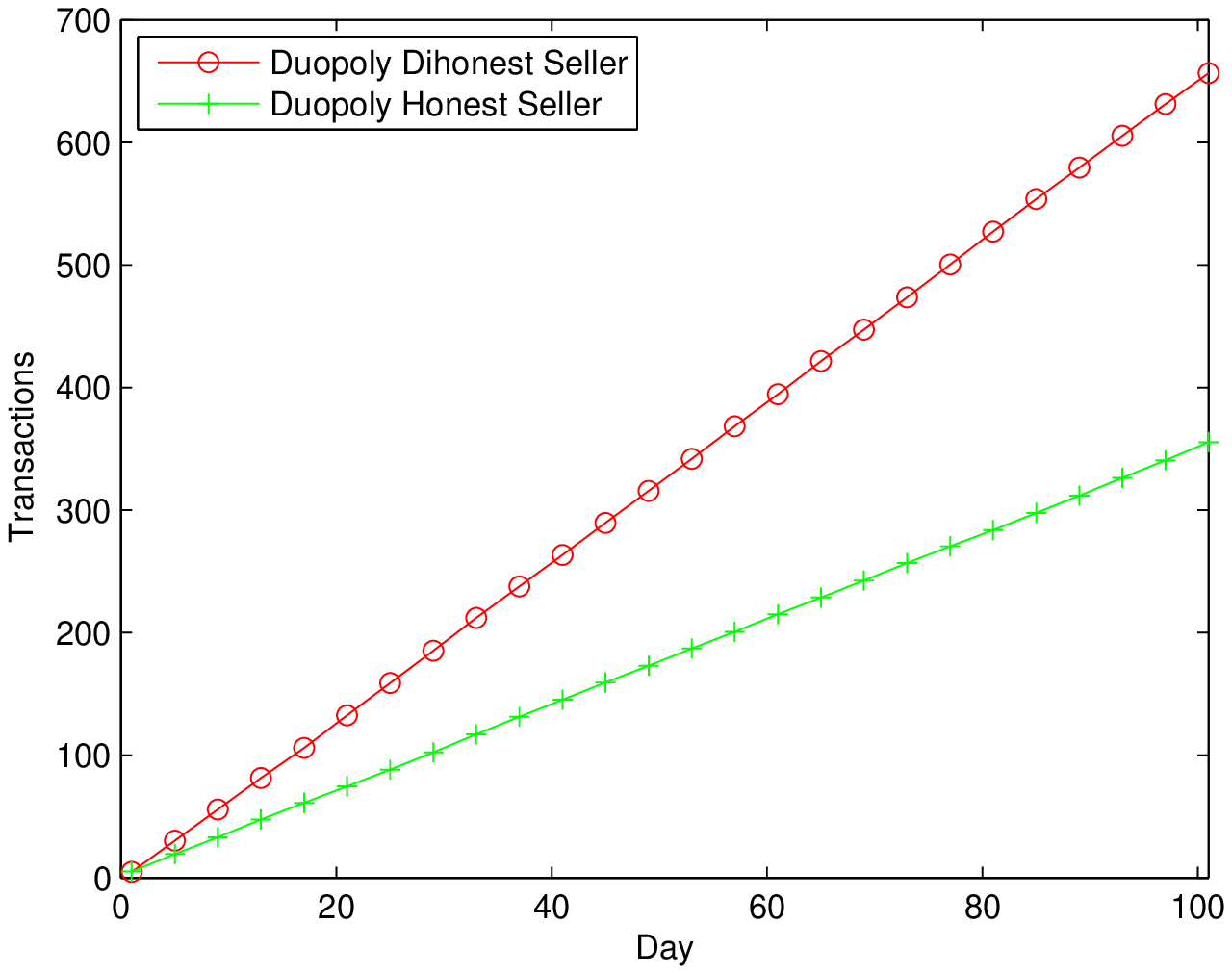}
\caption{BRS + TRAVOS vs. Sybil Attack}
\label{fig:bt_sybil}
\end{figure}

\begin{figure}[!t]
\centering
\includegraphics[scale=0.7]{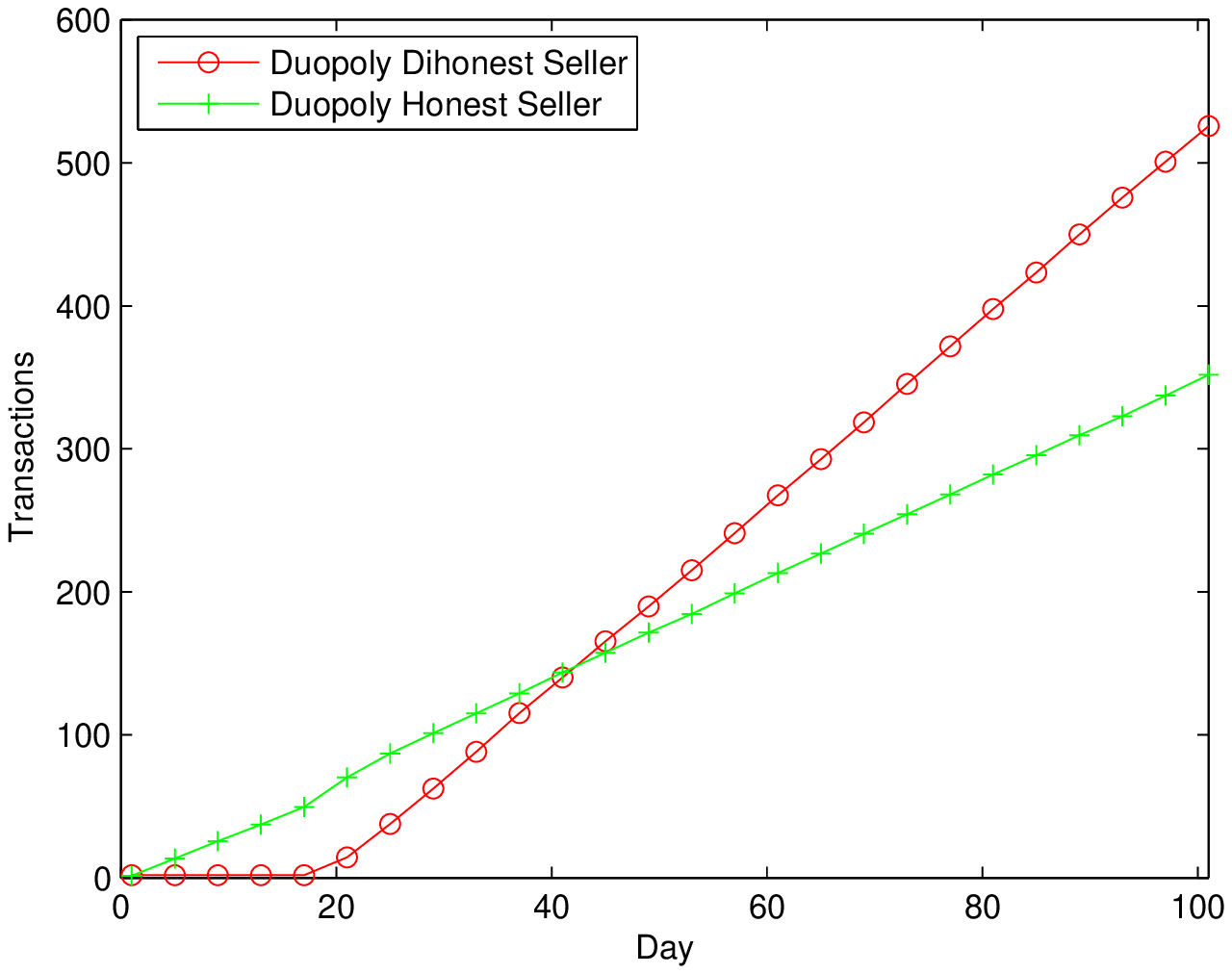}
\caption{BRS + TRAVOS vs. Sybil Camouflage Attack}
\label{fig:bt_sybil_camouflage}
\end{figure}

\begin{figure}[!t]
\centering
\includegraphics[scale=0.7]{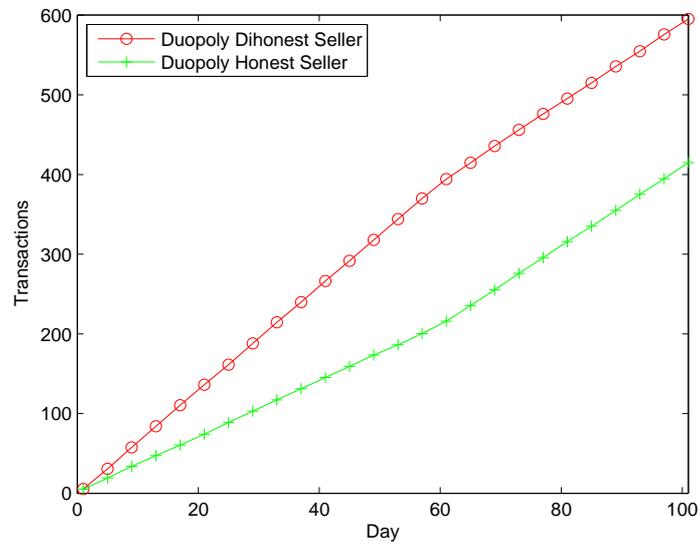}
\caption{BRS + TRAVOS vs. Sybil Whitewashing Attack}
\label{fig:bt_sybil_whitewashing}
\end{figure}

\section{Robustness Evaluation}

Eight possible combinations of trust models are obtained and their robustness against all the attacks have been evaluated. Notice that the new model name follows the order of using the two different models.
For instance, BRS + TRAVOS means using BRS to filter out unfair ratings then using TRAVOS to discount the remaining ratings in the evaluation of the sellers' reputation.

We will discuss the robustness enhancement of each combined model against all attacks based on the experimental results presented in Table~\ref{table:RobustnessCombine}.

\subsection{Filter-then-Discount}

\subsubsection{BRS + TRAVOS and BRS + Personalized}
Similar to BRS, they are still vulnerable to many attacks such as Whitewashing Attack, Sybil Attack, Sybil Camouflage Attack, and Sybil Whitewashing Attack. The reason is, under these attacks BRS will inaccurately filter out some honest buyers' ratings and keep some dishonest buyers' ratings after the first step of Approach 1; the remaining unfair ratings will be used by Discounting-based trust models to inaccurately suggest honest buyers to transact with the dishonest duopoly seller.

Fig.~\ref{fig:bt_whitewashing}---Fig.~\ref{fig:bt_sybil_whitewashing} depict under Whitewashing Attack, Sybil Attack, Sybil Camouflage Attack, and Sybil Whitewashing Attack, how the transactions of the duopoly sellers grow day after day when BRS + TRAVOS is used by honest buyers to decide which duopoly seller to transact with. The negative transaction volume difference between the honest and dishonest duopoly seller on Day 100 indicates that BRS + TRAVOS is vulnerable to these attacks.

\subsubsection{iCLUB + TRAVOS and iCLUB + Personalized}
\begin{figure}[!t]
\centering
\includegraphics[scale=0.7]{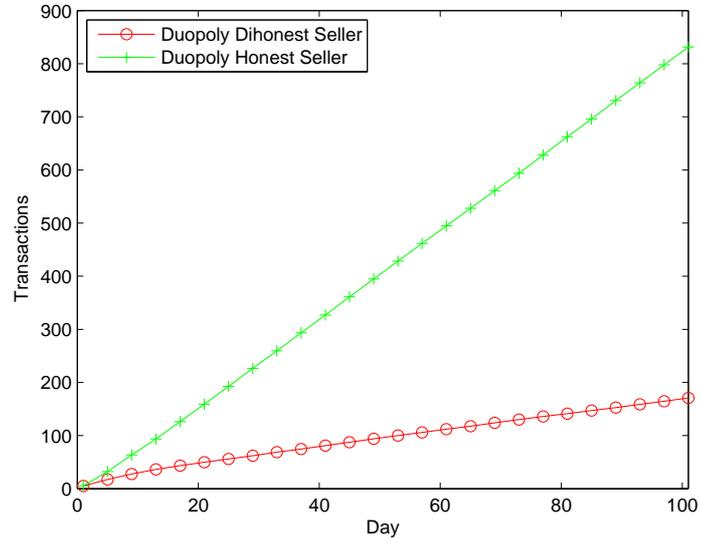}
\caption{iCLUB + Personalized vs. Whitewashing Attack}
\label{fig:ip_whitewashing}
\end{figure}

\begin{figure}[!t]
\centering
\includegraphics[scale=0.7]{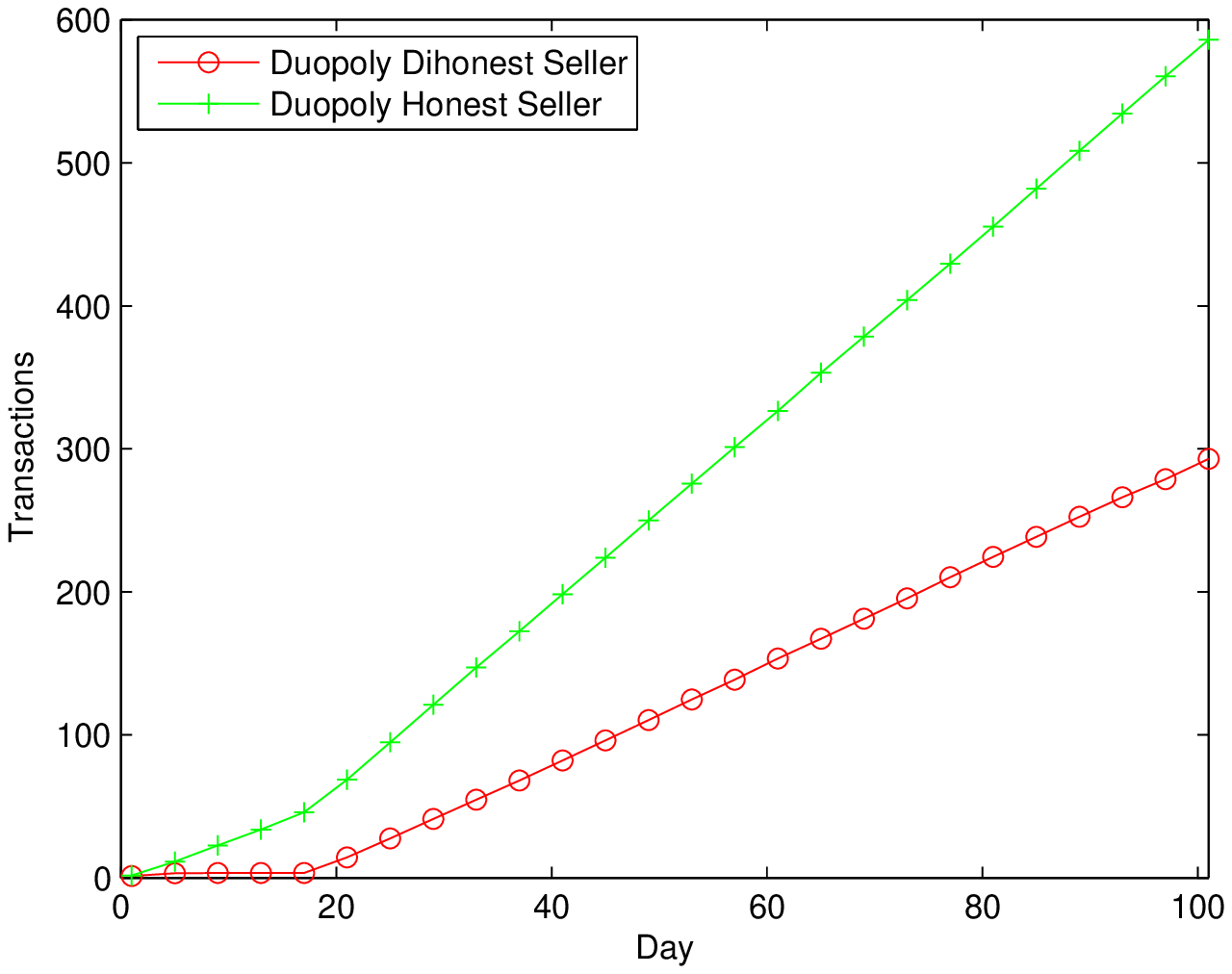}
\caption{iCLUB + Personalized vs. Sybil Camouflage Attack}
\label{fig:ip_sybil_camouflage}
\end{figure}

\begin{figure}[!t]
\centering
\includegraphics[scale=0.7]{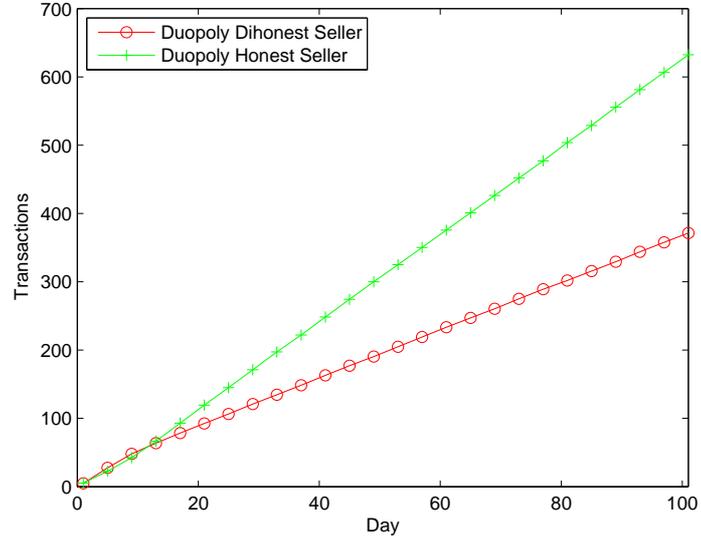}
\caption{iCLUB + Personalized vs. Sybil Attack}
\label{fig:ip_sybil}
\end{figure}

\begin{figure}[!t]
\centering
\includegraphics[scale=0.7]{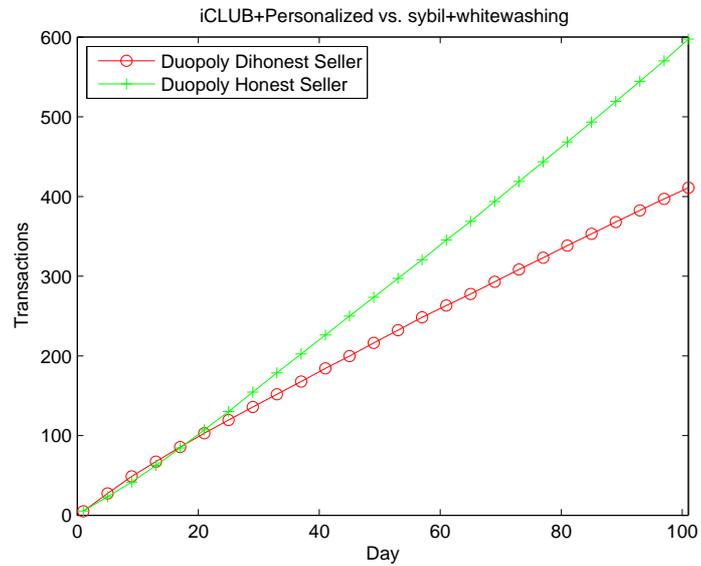}
\caption{iCLUB + Personalized vs. Sybil Whitewashing Attack}
\label{fig:ip_sybil_whitewashing}
\end{figure}

Contrary to BRS, iCLUB is robust against Whitewashing Attack and Sybil Camouflage Attack. Therefore, iCLUB + TRAVOS and iCLUB + Personalized are also able to effectively filter out unfair ratings at the first step of Approach 1, and are robust against these attacks (Fig.~\ref{fig:ip_whitewashing} and Fig.~\ref{fig:ip_sybil_camouflage}). However, due to the instability of the robustness of iCLUB against Sybil Attack and Sybil Whitewashing Attack, iCLUB + TRAVOS and iCLUB + Personalized are still not completely robust against these attacks (Fig.~\ref{fig:ip_sybil} and Fig.~\ref{fig:ip_sybil_whitewashing}).

\subsection{Discount-then-Filter}
The complete robustness of TRAVOS and Personalized against Whitewashing Attack ensures all the attackers' ratings will be filtered out at the first step of Approach 2.

As described in Chapter~\ref{section:RobustnessOfSingleTrustModels}, although TRAVOS and Personalized are unable to discount the trustworthiness of a Sybil, Sybil Camouflage or Sybil Whitewashing Attacker to a large extent (soft punishment: only slightly lower than 0.5), the threshold value we choose ($\epsilon = 0.5$) is able to filter out all these attackers' ratings at the second step of Approach 2. Therefore, Personalized + BRS and Personalized + iCLUB are completely robust against Sybil Attack, Sybil Camouflage Attack and Sybil Whitewashing Attack. Likewise, TRAVOS + BRS and TRAVOS + iCLUB are completely robust against most attacks.

\begin{figure}[!t]
\centering
\includegraphics[scale=0.7]{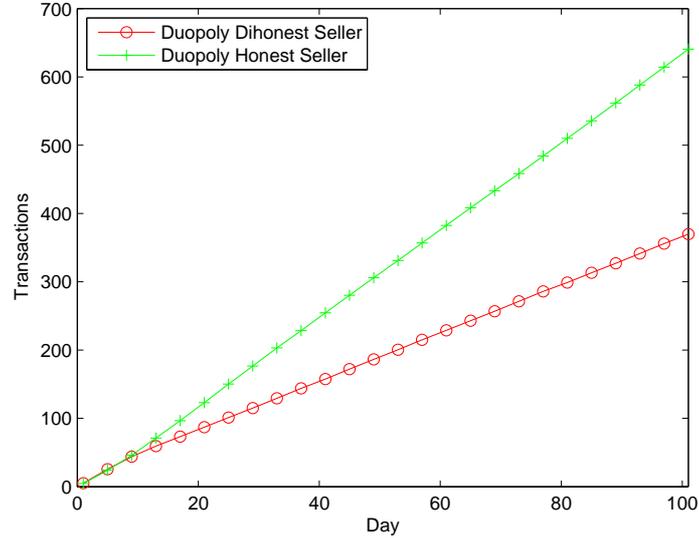}
\caption{Personalized + iCLUB vs. Sybil Attack}
\label{fig:pi_sybil}
\end{figure}

\begin{figure}[!t]
\centering
\includegraphics[scale=0.7]{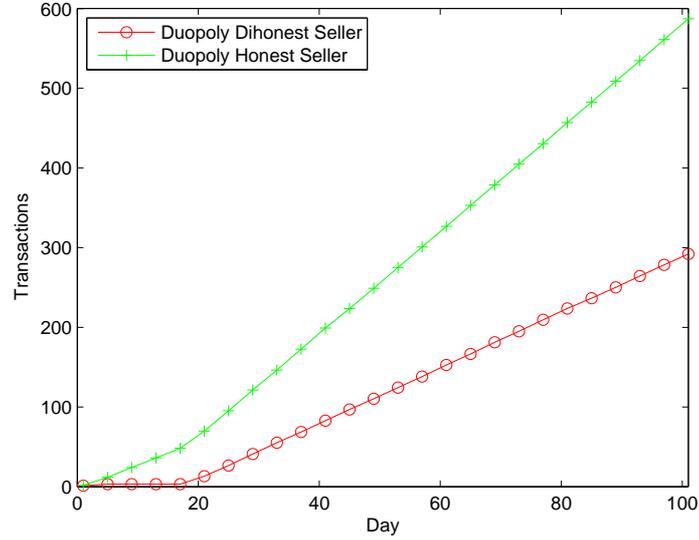}
\caption{Personalized + iCLUB vs. Sybil Camouflage Attack}
\label{fig:pi_sybil_camouflage}
\end{figure}

\begin{figure}[!t]
\centering
\includegraphics[scale=0.7]{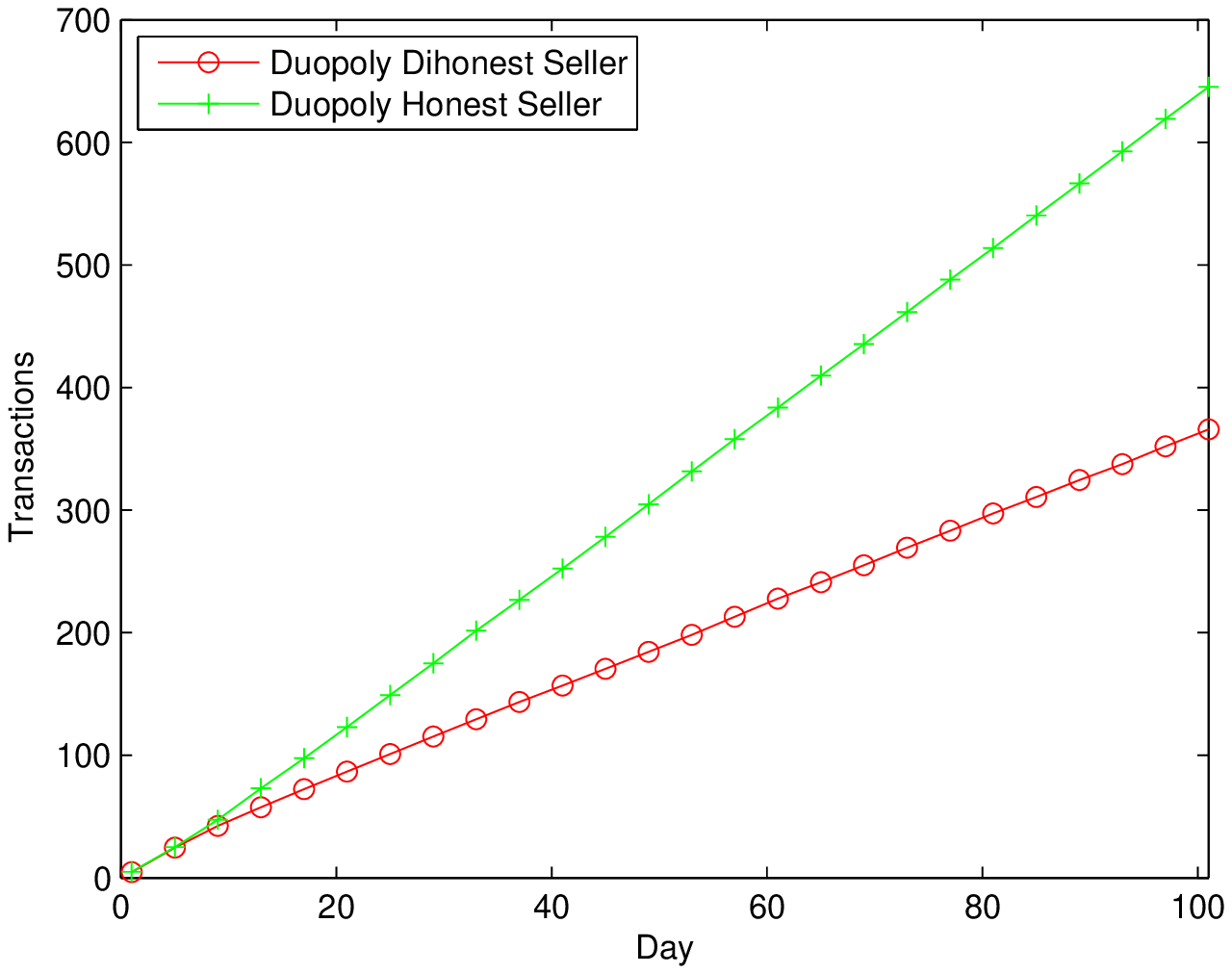}
\caption{Personalized + iCLUB vs. Sybil Whitewashing Attack}
\label{fig:pi_sybil_whitewashing}
\end{figure}

Fig.~\ref{fig:pi_sybil}---Fig.~\ref{fig:pi_sybil_whitewashing} show the complete robustness of Personalized + iCLUB against Sybil Attack, Sybil Camouflage Attack and Sybil Whitewashing Attack.

\begin{figure}[!t]
\centering
\includegraphics[scale=0.7]{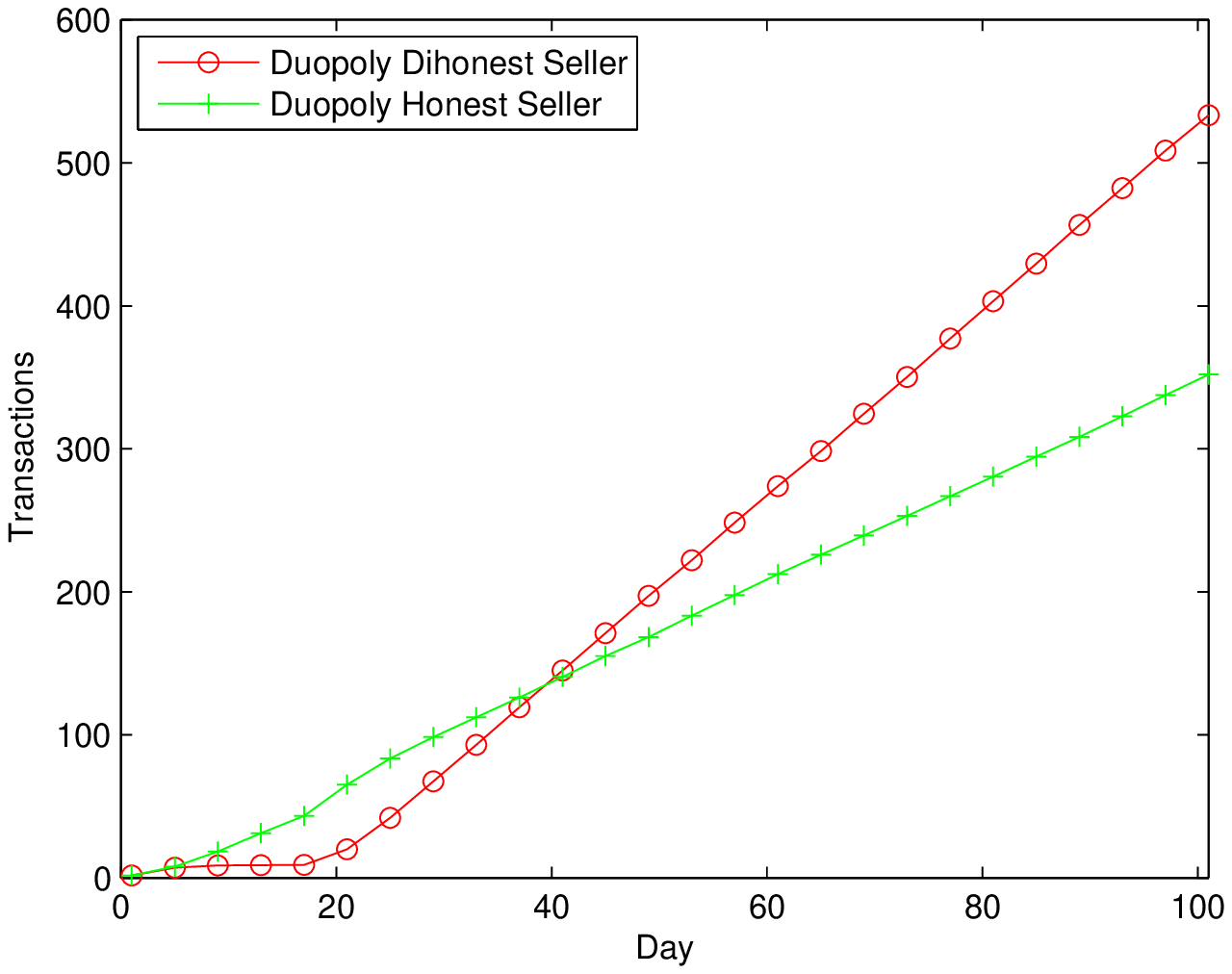}
\caption{TRAVOS + BRS vs. Sybil Camouflage Attack}
\label{fig:tb_sybil_camouflage}
\end{figure}

One exception is that, TRAVOS + BRS is still vulnerable to Sybil Camouflage Attack (Fig.~\ref{fig:tb_sybil_camouflage}). This is because TRAVOS inaccurately promotes attackers' trustworthiness (most are slightly higher than 0.5) and their ratings are unable to be filtered out at the second step of Approach 2. Unlike iCLUB, which is robust against Sybil Camouflage Attack, BRS is vulnerable to it.

\begin{figure}[t!]
\centering
\includegraphics[scale=0.7]{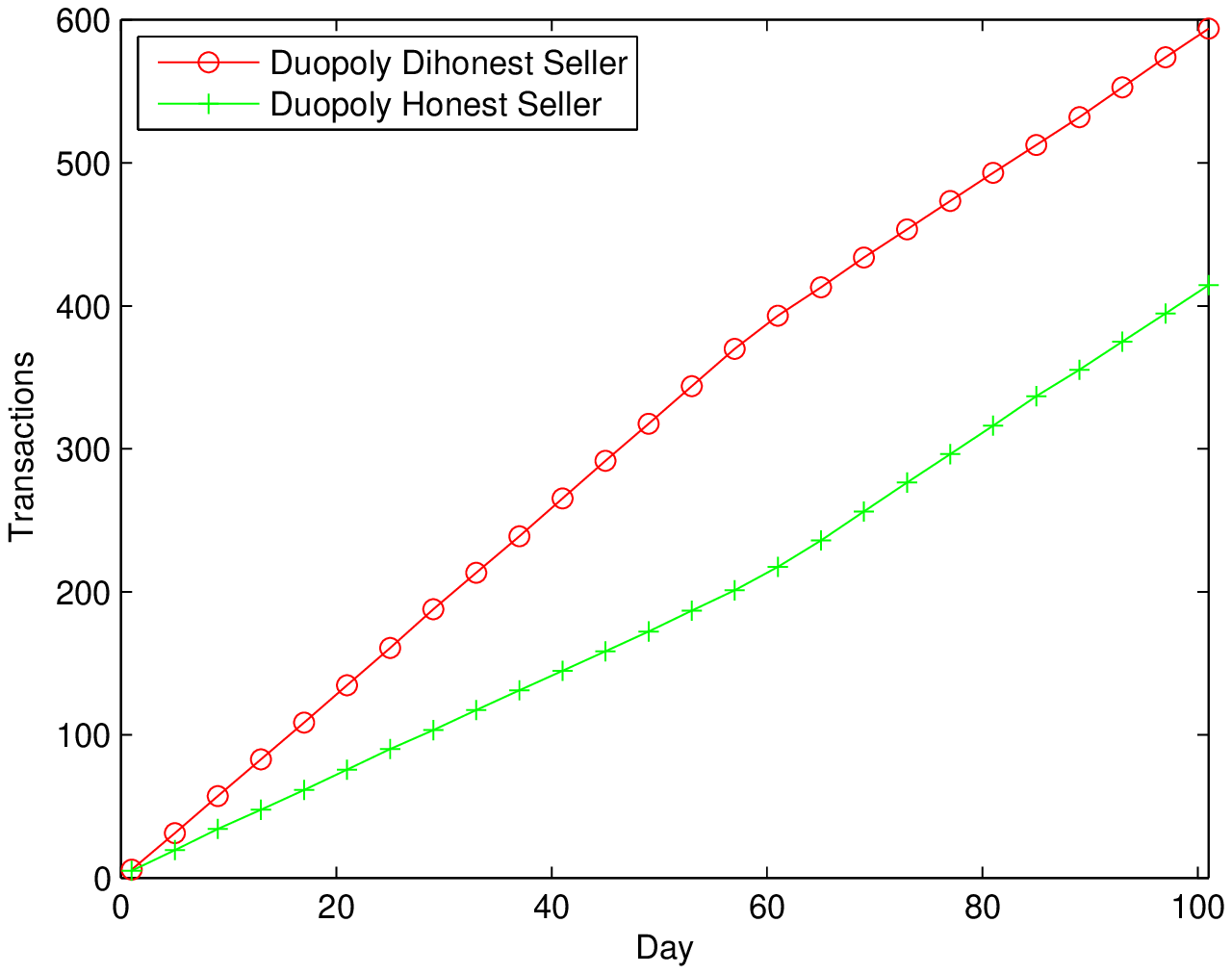}
\caption{BRS vs. Sybil Whitewashing Attack}
\label{fig:brs_sw}
\end{figure}

\begin{figure}[t!]
\centering
\includegraphics[scale=0.7]{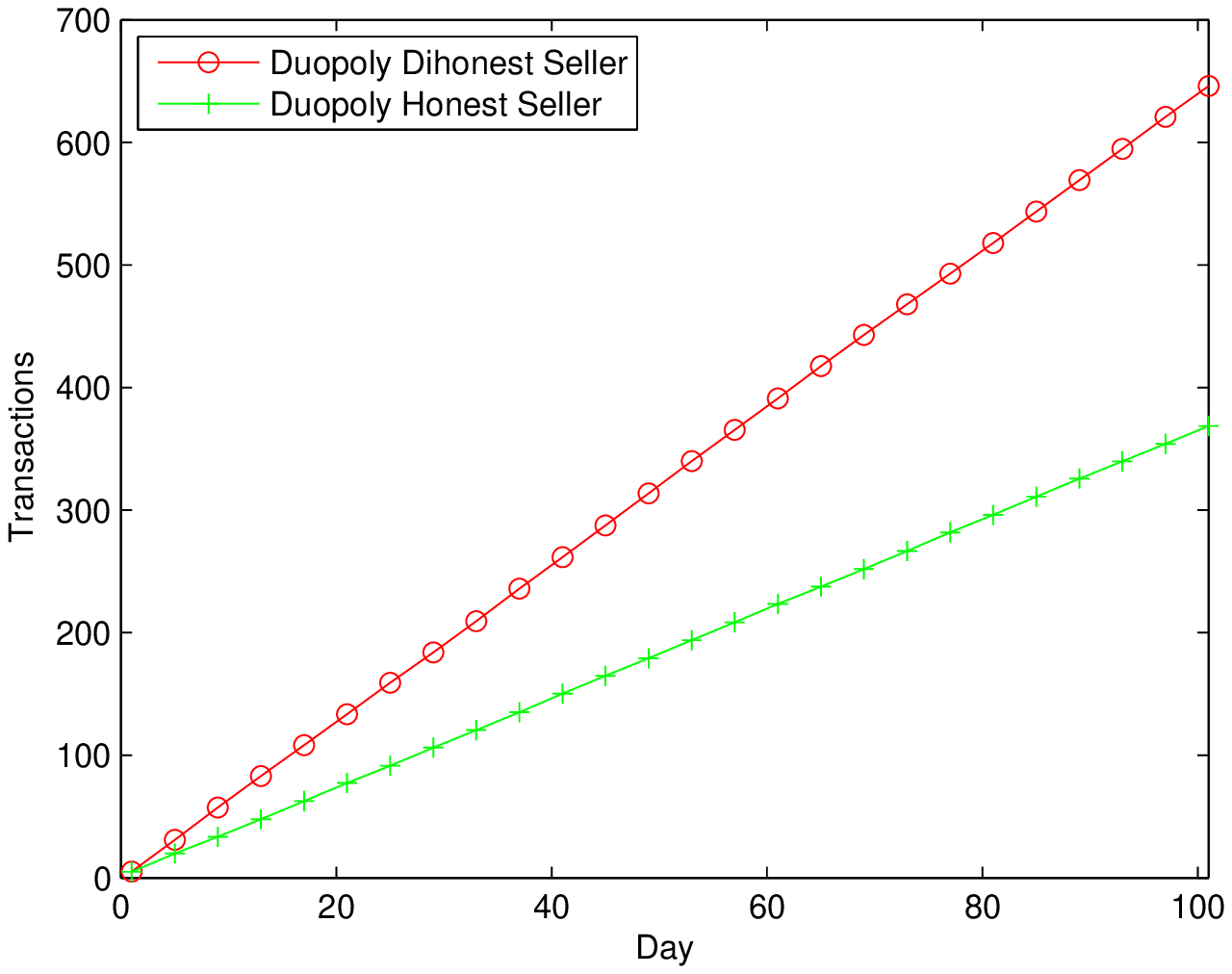}
\caption{Personalized vs. Sybil Whitewashing Attack}
\label{fig:personalized_sw}
\end{figure}

\begin{figure}[t!]
\centering
\includegraphics[scale=0.7]{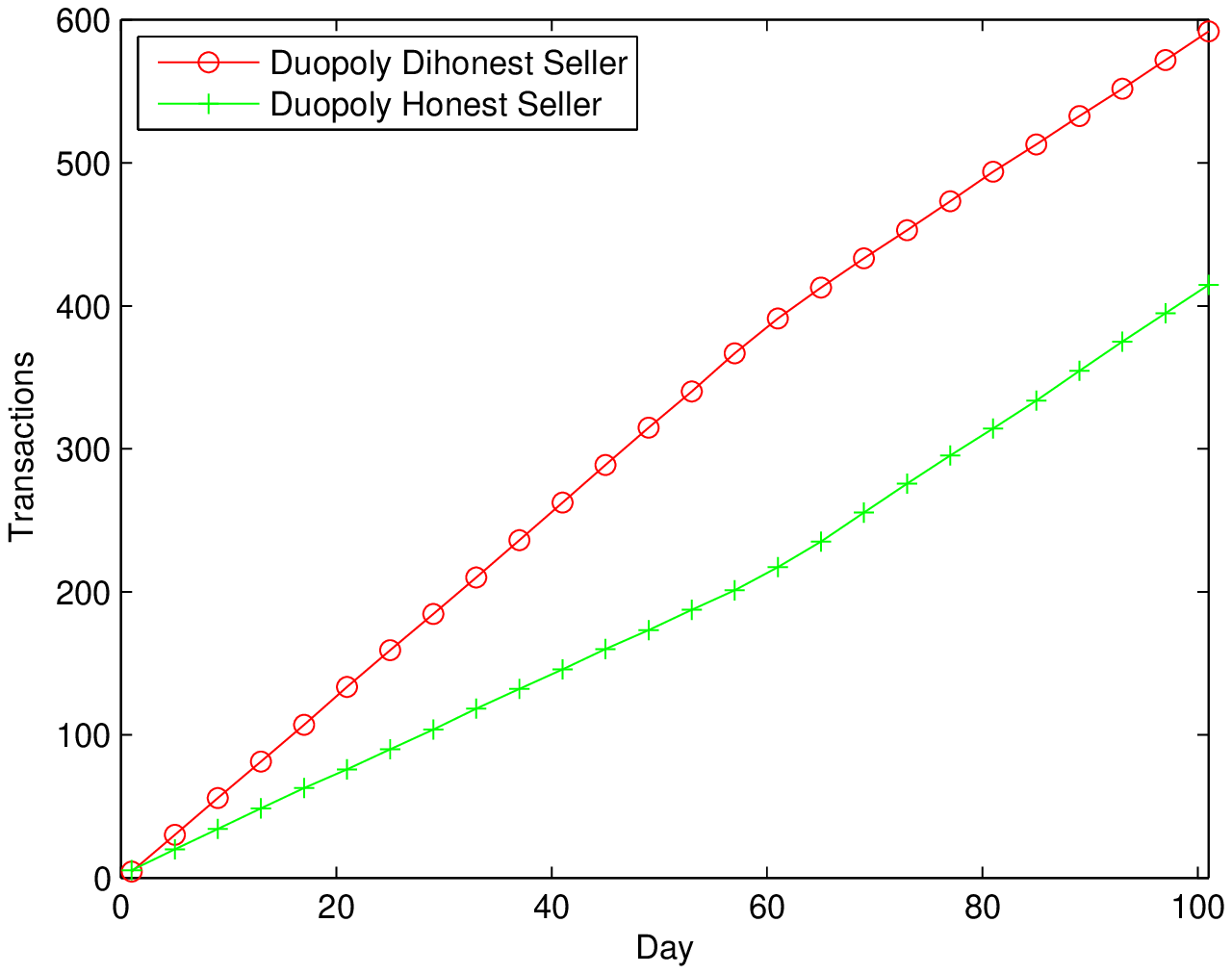}
\caption{BRS + Personalized vs. Sybil Whitewashing Attack}
\label{fig:bp_sw}
\end{figure}

\begin{figure}[t!]
\centering
\includegraphics[scale=0.7]{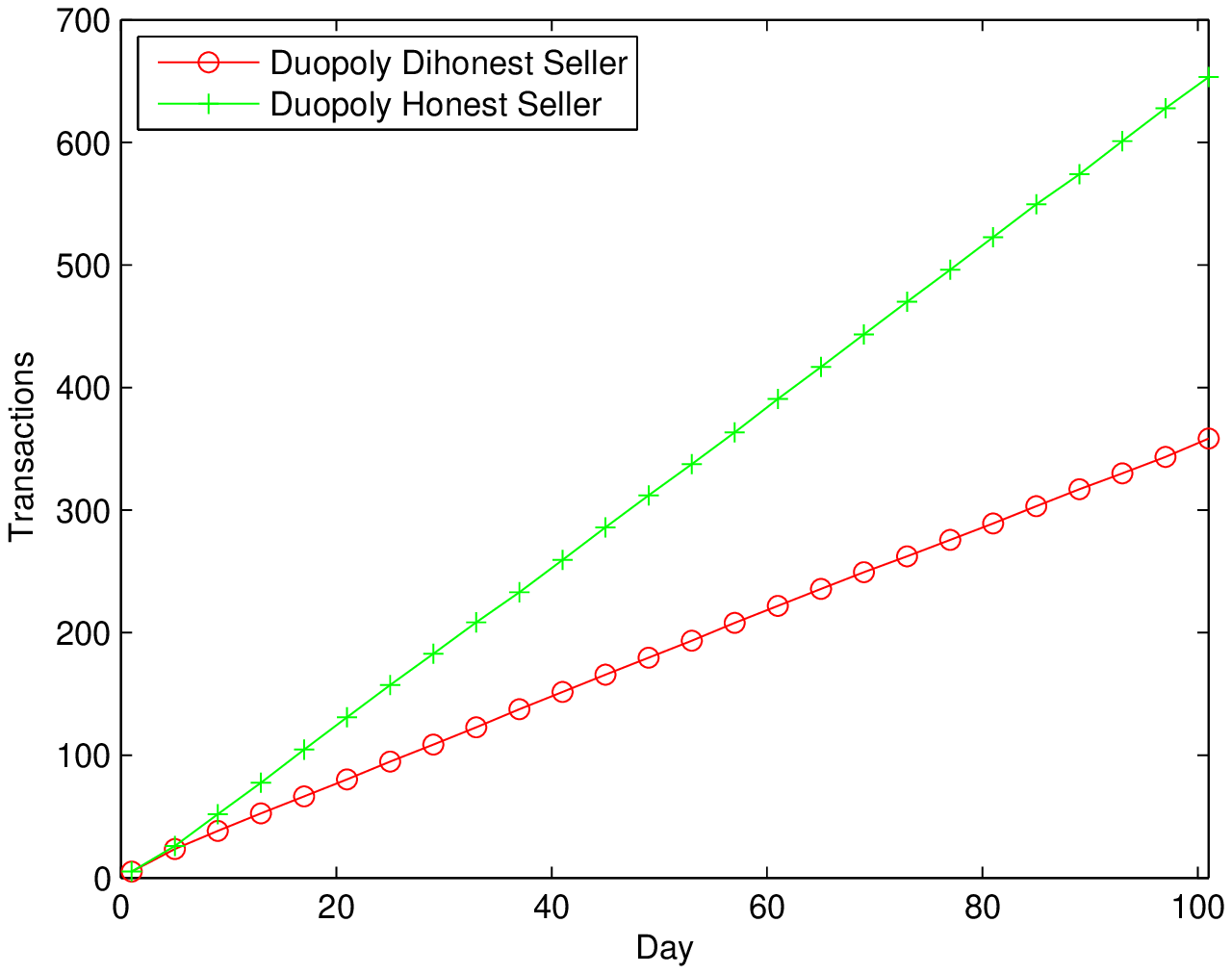}
\caption{Personalized + BRS vs. Sybil Whitewashing Attack}
\label{fig:pb_sw}
\end{figure}

\subsection{Conclusions}
Based on the results in Table~\ref{table:RobustnessSingle} and Table~\ref{table:RobustnessCombine}, we conclude that, robustness of single trust models can be enhanced by combining different categories, and Discount-then-Filter is most robust. Particularly, TRAVOS + iCLUB, Personalized + BRS, and Personalized + iCLUB are robust against all the investigated attacks.

Fig.~\ref{fig:brs_sw}---Fig.~\ref{fig:pb_sw} show how the robustness of the trust models is enhanced with the Discount-then-Filter approach, while Filter-then-Discount is still vulnerable. In other words, if the e-marketplace is equipped with either BRS or Personalized, the duopoly dishonest seller is able to gain a higher transaction volume than that of the duopoly honest seller by hiring or collaborating with the Sybil Whitewashing attackers. Therefore, as the time goes, this e-marketplace will be filled with more and more dishonest sellers until it fails with all the honest buyers exiting the market. In contrast, with Personalized + BRS, the Discount-then-Filter combined trust model, only by behaving honestly is the duopoly seller able to gain a higher transaction volume; thus, sellers are less motivated to hire or collaborate with advisors providing unfair ratings. In this way, e-commerce is better safeguarded against unfair ratings---the advisor cheating behaviors.



\def\baselinestretch{1}
\chapter{Conclusion and Future Work}
\label{section:Conclusion}
\ifpdf
    \graphicspath{{Conclusions/ConclusionsFigs/PNG/}{Conclusions/ConclusionsFigs/PDF/}{Conclusions/ConclusionsFigs/}}
\else
    \graphicspath{{Conclusions/ConclusionsFigs/EPS/}{Conclusions/ConclusionsFigs/}}
\fi

\def\baselinestretch{1.66}

\section{Conclusion}
\markboth{\MakeUppercase{\thechapter. Conclusion and Future Work}}{\thechapter. Conclusion and Future Work}
Trust models can benefit us in choosing trustworthy sellers to transact with in the e-marketplace only when they are robust against external unfair rating attacks. Recently it is argued some trust models are vulnerable to certain attacks and they are not as robust as what their designers claimed to be. Therefore, robustness of trust models for handling unfair ratings have to be evaluated under a comprehensive attack environment to make the results more credible.

In this project, we designed an extendable e-marketplace testbed to incorporate each existing trust model under a comprehensive set of attack models to evaluate the robustness of trust models. To the best of our knowledge, this is the first demonstration that multiple vulnerabilities of trust models for handling unfair ratings do exist. We conclude that, in our experiments there is no single trust model that is robust against all the investigated attacks. While we have selected a small number of trust models for this initial study, we can hardly believe that other trust model will not have these vulnerabilities.
We argue that, in the future any newly proposed trust model at least has to demonstrate robustness (or even complete robustness) to these attacks before being claimed as effective in handling unfair ratings.

To address the challenge of the existing trust models' multiple vulnerabilities, we classified the existing trust models into two categories: Filtering-based and Discounting-based, and further proposed two approaches to combining the existing trust models from different categories: Filter-then-Discount and Discount-then-Filter. We for the first time proved that most of the Discount-then-Filter combinations are robust against all the investigated attacks. With such combined trust models, only by behaving honestly are sellers able to gain higher transaction volumes; thus, sellers are less motivated to hire or collaborate with advisors providing unfair ratings. In this way, e-commerce is better safeguarded against unfair ratings---the advisor cheating behaviors.

A concise version of this report titled ``Robustness of Trust Models and Combinations for Handling Unfair Ratings" was accepted by the \emph{6th IFIP WG 11.11 International Conference on Trust Management (IFIPTM'12)} (\cite{zhang2012robustness}).

\section{Future Work}
Although our work focused on unfair rating attacks, we plan to combine sellers' cheating behaviors with advisors' unfair ratings, and evaluate their threats to the existing trust models. We are also interested in re-designing new trust models to be completely robust against all the investigated attacks without combining existing ones. Since Sybil-based unfair ratings attacks are more effective than Non-Sybil-based, we also want to design more effective unfair rating attacks with limited buyer account resources.

We believe these directions inspired by this work will yield further important insights in the trust management area.




\appendix

\bibliographystyle{plainnat}
\renewcommand{\bibname}{References} 
\bibliography{References/references} 

\end{document}